\documentclass[aps,prl,preprint,superscriptaddress]{revtex4-1}
\usepackage{graphicx}
\usepackage{color}
\usepackage{amsfonts}
\usepackage{isomath}
\usepackage{amsmath}
\usepackage{amsthm}
\usepackage{amsfonts}
\usepackage{braket}

\newcommand{\EF}{$E_\mathrm{F}$}

\usepackage{yhmath}

\def\WS2{WS$_2$}
\def\MoS2{MoS$_2$}{

\def\EF{$E\mathrm{_F}$}

\def\degrees{$^{\circ}$}

\renewcommand{\vec}[1]{\mathbfit{#1}}

\usepackage{setspace}

\begin{document}

\title{Nanoscale view of engineered massive Dirac quasiparticles in lithographic superstructures}

\author{Alfred J. H. Jones}
\affiliation{Department of Physics and Astronomy, Aarhus University, 8000 Aarhus C, Denmark}
\author{Lene Gammelgaard}
\affiliation{DTU Physics, Technical University of Denmark, 2800 Kgs. Lyngby, Denmark}
\affiliation{Center for Nanostructured Graphene, Technical University of Denmark, 2800 Kgs. Lyngby, Denmark}
\author{Mikkel O. Sauer}
\affiliation{Department of Materials and Production, Aalborg University, 9220 Aalborg Øst, Denmark}
\affiliation{Department of Mathematical Science, Aalborg University, 9220 Aalborg Øst, Denmark}
\affiliation{Center for Nanostructured Graphene (CNG), 9220 Aalborg Øst, Denmark}
\author{Deepnarayan Biswas}
\affiliation{Department of Physics and Astronomy, Aarhus University, 8000 Aarhus C, Denmark}
\author{Roland J. Koch}
\author{Chris Jozwiak}
\author{Eli Rotenberg}
\author{Aaron Bostwick}
\affiliation{Advanced Light Source, E. O. Lawrence Berkeley National Laboratory, Berkeley, California 94720, USA}
\author{Kenji~Watanabe}
\affiliation{Research Center for Functional Materials, 
	National Institute for Materials Science, 1-1 Namiki, Tsukuba 305-0044, Japan}
\author{Takashi~Taniguchi}
\affiliation{International Center for Materials Nanoarchitectonics, 
	National Institute for Materials Science,  1-1 Namiki, Tsukuba 305-0044, Japan}
\author{Cory R. Dean}
\affiliation{Department of Physics, Columbia University, New York, NY, USA}
\author{Antti-Pekka Jauho}
\author{Peter B{\o}ggild}
\affiliation{DTU Physics, Technical University of Denmark, 2800 Kgs. Lyngby, Denmark}
\affiliation{Center for Nanostructured Graphene, Technical University of Denmark, 2800 Kgs. Lyngby, Denmark}
\author{Thomas G. Pedersen}
\affiliation{Department of Materials and Production, Aalborg University, 9220 Aalborg Øst, Denmark}
\affiliation{Center for Nanostructured Graphene (CNG), 9220 Aalborg Øst, Denmark}
\author{Bjarke S. Jessen}
\email{bjarke.jessen@columbia.edu}
\affiliation{Department of Physics, Columbia University, New York, NY, USA}
\author{S{\o}ren Ulstrup}
\email{ulstrup@phys.au.dk}
\affiliation{Department of Physics and Astronomy, Aarhus University, 8000 Aarhus C, Denmark}

\begin{abstract}
Massive Dirac fermions are low-energy electronic excitations characterized by a hyperbolic band dispersion. They play a central role in several emerging physical phenomena such as topological phase transitions, anomalous Hall effects and superconductivity.  This work demonstrates that massive Dirac fermions can be controllably induced by lithographically patterning superstructures of nanoscale holes in a graphene device.  Their band dispersion is systematically visualized using angle-resolved photoemission spectroscopy with nanoscale spatial resolution.  A linear scaling of effective mass with feature sizes is discovered, underlining the Dirac nature of the superstructures.  \textit{In situ} electrostatic doping dramatically enhances the effective hole mass and leads to the direct observation of an electronic band gap that results in a peak-to-peak band separation of (0.64 $\pm$ 0.03)~eV, which is shown via first-principles calculations to be strongly renormalized by carrier-induced screening.  The presented methodology outlines a new approach for band structure engineering guided by directly viewing structurally- and electrically-tunable massive Dirac quasiparticles in lithographic superstructures at the nanoscale.
\end{abstract}

\maketitle

\section{Introduction}
Superlattices are transformative for electronic properties of two-dimensional (2D) materials, as exemplified by the realization of massive Dirac fermions, quantum fractal states, and unconventional ferroelectricity using lattice mismatched graphene on hexagonal boron nitride (hBN) substrates \cite{Dean:2013,Hunt2013, Zheng:2020} and by achieving strongly correlated states in magic angle twisted bilayer graphene \cite{Cao2:2018,Cao2018}. Complex surface interactions can be engineered using superlattice potentials, allowing for selective adsorption of hydrogen atoms in graphene moir{\'e}s on lattice mismatched metallic substrates, which has been shown to open a gap in the Dirac cone \cite{Balog2010}. However, such moir\'e-derived effects are limited by the lattices of the constituent monolayers, as well as disorder from spatially varying twist-angles \cite{Uri:2020}. Lithographic top-down patterning of nanoscale superstructures can offer significant improvements in flexibility and repeatability compared to moir\'e superlattices, as smaller selected regions can be arbitrarily structured in graphene with feature sizes down to tens of nanometers \cite{Berger2006,Han:2007,Ponomarenko2008,Eroms2009}. Inspired by the observation of anomalous Hall conduction in semiconductor heterostructures with periodic patterns of holes, so-called \textit{antidot lattices} \cite{Weiss:1991}, there have been several theoretical proposals of engineering a mass gap in graphene by top-down patterning a periodic arrangement of holes in the graphene sheet   \cite{Pedersen2008,Furst2009,Brun2014}. These structures may now be realized with extremely high precision and edge quality using lithographic techniques \cite{Danielsen:2021,Jessen2019}, thereby enabling new experiments to investigate the nature of massive Dirac fermion quasiparticles and associated band gaps emerging in such systems. 

\section{Results and Discussion}

\begin{figure*} [t!]
	\begin{center}
		\includegraphics[width=1\textwidth]{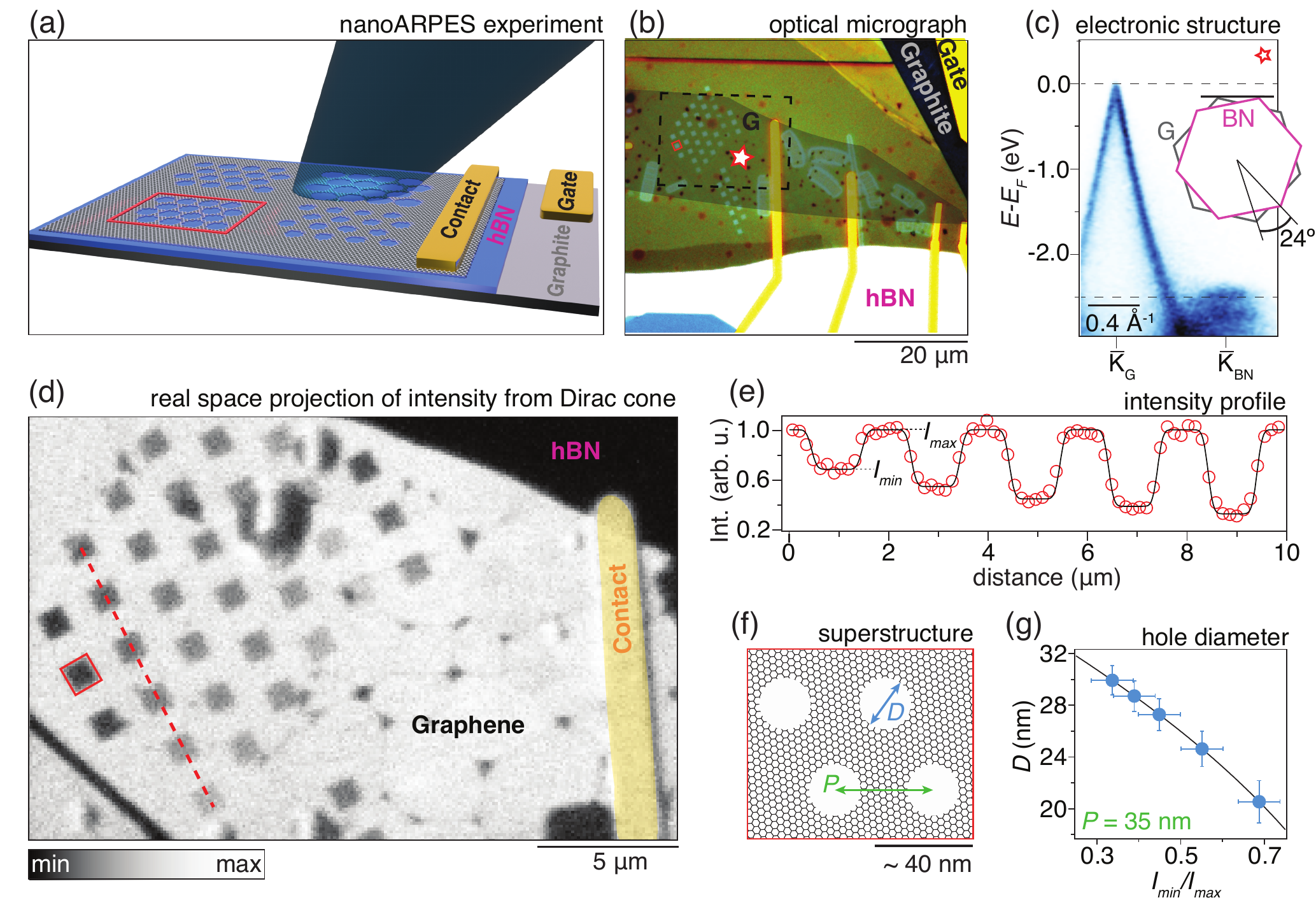}
		\caption{Nanoscale photoemission experiment on patterned graphene device.  a) Illustration of experimental setup.  b) Optical micrograph of the device with shape and position of the graphene flake outlined.  c) Electronic band structure measured from the pristine graphene region indicated by a star in (b).  The twist angle between graphene and the underlying hBN is indicated on the BZs shown in the inset.  d) Real space map of the photoemission intensity integrated over the $(E,k)$-region containing the Dirac cone between the dashed horizontal lines in (c). The measured area is demarcated by a dashed rectangle in (b).  e) Intensity profile extracted along the dashed red line in (d), resulting in alternating intensity minima ($I_{min}$) and maxima ($I_{max}$) from patterned and pristine graphene areas, respectively. The solid curve is a fit to a square wave convoluted by the spatial resolution transfer function.  f) Diagram of the superstructure resulting from lithographic hole patterns with hole diameter $D$ and periodicity $P$.  g) Extracted hole diameter (markers) estimated from the mean intensity ratios fitted in (e).  Error bars on the intensity ratios represent one standard deviation, as given by the fit in (e).  Error bars on $D$ have been propagated from the intensity ratios. The curve is a plot of the geometric relation between  $D$ and the intensity ratio (see text). The lithographically defined periodicity is 35 nm for the patterns analysed in (e) and (g).}
		\label{fig:1}
	\end{center}
\end{figure*}

The electronic structure of a nano-engineered device consisting of a periodic pattern of holes defined in a graphene flake by electron-beam lithography (see details in Experimental Section and Figure \ref{fig:DevFig})  is measured using angle-resolved photoemission spectroscopy with nanoscale spatial resolution (nanoARPES).  A synchrotron light spot is focused on the order of 200 nm on the sample using precision piezo-mechanically controlled Fresnel zone plate optics  \cite{Kastl:2019}. An illustration of the experimental setup is presented in \textbf{Figure \ref{fig:1}}a. The nanoscale light spot is scanned across multiple patterned regions within the same single-crystal graphene flake in order to photoemit electrons and measure the position-resolved energy- and momentum-dependent electronic dispersion. Neighbouring patterned regions are separated by an area of pristine graphene, providing a clean internal reference for the measurements on patterned regions. The patterned flake is positioned on a graphite back-gate with an hBN dielectric separation layer, thereby realizing a charge carrier-tunable device that is compatible with nanoARPES measurements \cite{Nguyen:2019,Muzzio2020}. The device is shown in the optical micrograph in Figure \ref{fig:1}b, with graphene, hBN and graphite flakes labelled. Several square and rectangular regions are visible in the graphene flake, resulting from lithographic etching. We will focus on the square regions in the discussion below. 

The electronic structure from a pristine part of the graphene flake, which is marked by a star in Figure \ref{fig:1}b, is shown in Figure \ref{fig:1}c. The dispersion is measured along the $\bar{\mathrm{K}}_\mathrm{G} - \bar{\mathrm{K}}_\mathrm{BN}$ direction of the graphene and hBN Brillouin zones (BZs), as sketched in the inset. The Dirac cone of graphene is visible in addition to the valence band maximum of the underlying hBN at -2.5 eV. The momentum-separation between these points corresponds to a twist angle of $(24 \pm 0.5)$\degrees, as shown in Figure \ref{fig:S1}. The large angle is advantageous for our study, as we can disregard moir{\'e} effects stemming from the graphene-hBN interaction \cite{Wang:2016, Jessen2019}.

We proceed to identify the electronic properties of the patterned regions by collecting the four-dimensional $(E,k,x,y)$-dependent photoemission intensity over the area demarcated by dashed lines in Figure \ref{fig:1}b. The $(E,k)$-window of the measurement corresponds to that shown in Figure \ref{fig:1}c. The measured intensity of the Dirac cone is then energy- and momentum-integrated between the dashed horizontal lines in Figure \ref{fig:1}c and projected onto the real space $(x,y)$-coordinates of the nanoARPES scan, leading to the spatial distribution of Dirac cone intensity shown in Figure \ref{fig:1}d. The bright parts of the map derive from pristine graphene with the electronic structure in Figure \ref{fig:1}c. The dark squares are  $1 \times 1$~$\mu$m$^2$ areas, each containing a different periodic pattern of etched holes. Further aspects of the spatial map, including the assignment of hBN and contact areas, are explored in Figure \ref{fig:S2}. 

By design, the pattern of holes adheres to a hexagonal unit cell characterized by periodicity $P$ and diameter $D$, shown schematically in Figure \ref{fig:1}f. Each column of patterned regions has been made with a fixed periodicity, incremented in steps of 5 nm between columns. Within each column the hole diameter is varied. However, while the spacing between patterned holes is known by design, the hole diameter sensitively depends upon the actual fabrication and etching conditions, which are inherently prone to variation on the nanometer scale \cite{Caridad:2018}. The diameter is therefore estimated from the nanoARPES map using the loss of intensity from graphene electronic states in the patterned regions. The intensity loss is demonstrated via the line-profile in Figure \ref{fig:1}e, which is extracted along the dashed red line in Figure \ref{fig:1}d. Increased removal of carbon atoms due to a larger diameter of patterned holes leads to a clear decreasing minimum intensity along the profile. The photoemission intensity of patterned and pristine areas are related to the hole diameter and periodicity via the proportion of removed graphene, given by $D = P\sqrt{(1-I_{min}/I_{max})(2\sqrt{3}/\pi)}$, which has been plotted in Figure \ref{fig:1}g. The extracted range of hole diameters is 21-30 nm,  in good agreement with the measured dimensions obtained for previous samples with scanning electron microscopy \cite{Mackenzie2017,Jessen2019}.

\begin{figure*} [ht!]
	\begin{center}
		\includegraphics[width=0.95\textwidth]{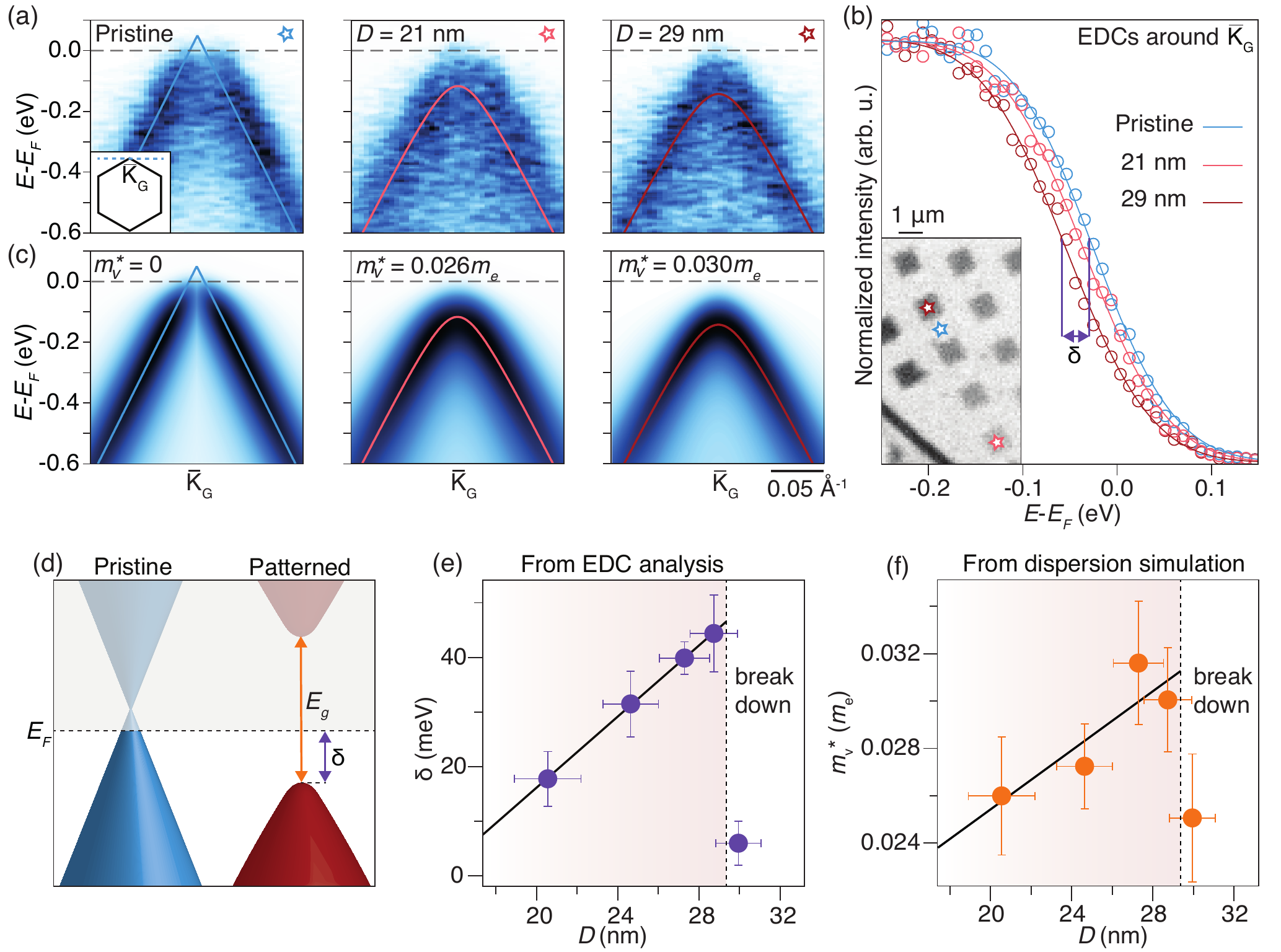}
		\caption{Observation of massive Dirac fermions.  a) ARPES intensity around $\bar{\mathrm{K}}_\mathrm{G}$ (see cut direction in BZ inset) for pristine and patterned graphene with the stated hole diameter.  b) EDCs (markers) with fits (curves) resulting from a momentum-integration of the ARPES intensity in (a).  Fits are composed of a Fermi-Dirac distribution multiplied by a linear background.  An EDC shift of the leading edge ($\delta$) to lower energy measured on the mid-point of the fitted distributions is indicated by a double-headed arrow.  The inset presents a spatial map with markers indicating the areas where the spectra are collected.  c) Simulations of the photoemission intensity optimized to the measurements shown in the same column in (a). The corresponding dispersion relations have been overlaid.  d)  Dispersions of massless and massive Dirac fermions, corresponding to the pristine and patterned graphene. The doping level is indicated by a dashed horizontal line. Purple and orange arrows indicate EDC shift ($\delta$) and band gap ($E_g$), respectively.  e) Mean values of $\delta$ extracted from the fits in (b) for patterns with increasing hole diameter.  Error bars on $\delta$ represent one standard deviation, as obtained from the fits in (b). f) Valence band effective mass ($m_v^{\ast}$) determined  by comparing the simulated intensity in (c) to the data in (a) and minimizing $\chi^2$. The error bars indicate one standard deviation for this minimization.  The overlaid lines in (e)-(f) represent linear fits that indicate antidot lattice scaling relations. The vertical dashed lines in (e) and (f) demarcate the breakdown of the scaling relations.}
		\label{fig:2}
	\end{center}
\end{figure*}

\textbf{Figure \ref{fig:2}}a presents detailed measurements of the ARPES intensity around the Fermi energy ($E_F$) comparing the $E(k)$-dispersion around $\bar{\mathrm{K}}_\mathrm{G}$ between pristine and patterned graphene areas (see areas marked by stars on the inset in Figure \ref{fig:2}b). The selected patterns have a periodicity of 35 nm and hole diameters of 21 nm and 29 nm, respectively. In pristine graphene, two linear branches of the Dirac cone are seen to disperse up to $E_F$ where their momentum-separation extrapolates to a Fermi wavevector ($k_F$) of $(0.015 \pm 0.002)$~\AA$^{-1}$, corresponding to hole-type ($p$) doping of $(0.7\pm 0.2)\cdot 10^{12}$~cm$^{-2}$. By contrast, the bands from the patterned areas exhibit a hyperbolic dispersion that is shifted down in energy with respect to $E_F$.  In order to quantify the magnitude of the shift we extract the intensity around the leading edge of energy distribution curves (EDCs) integrated over a range of  $\pm$0.05~\AA$^{-1}$ with respect to $\bar{\mathrm{K}}_\mathrm{G}$ (see details in Figure \ref{fig:S4}).  Such $k$-integrated EDCs have previously been demonstrated to provide a good description of the Fermi edge in graphene \cite{Ulstrup:2014a}. We therefore fit the EDCs using a Fermi-Dirac function multiplied by a linear background and extract the mid-point of the leading edge from the fit.  The data is shown in Figure \ref{fig:2}b with fits that reveal an increasing energy shift of the EDC leading edge ($\delta$) as the diameter of holes expands.  These shifts are interpreted as a quantitative measure of a band gap opening effect,  in line with the methodology commonly applied in ARPES measurements of gaps as established in studies of superconducting materials \cite{Damascelli:2003}.  

Spatially-dependent EDC shifts may also result from external charging effects caused by the underlying hBN or charged impurities that affect the patterned regions differently than the pristine regions of the graphene flake.  Electric fields caused by charging rigidly shift the kinetic energy distributions of all photoemitted electrons as they propagate in vacuum \cite{ULSTRUP2015340}.  EDCs of the background signal extracted at $k$ away from the graphene bands should therefore exhibit energy shifts that coincide with the values of $\delta$ extracted from the graphene bands.  We can rule out any energy shifts in the background as shown in Figure \ref{fig:S4}g,  excluding the possibility of such a spatially-dependent charging effect.

Charge carrier doping from possible remaining polymer residues and from impurities adsorbed on active edge sites in the holes of patterned graphene would also have an impact on the shape of the EDCs.  By analyzing the dispersion in 4 patches of pristine graphene surrounding the regions of patterned graphene, as shown in Figure \ref{fig:S4}, we find that the doping is homogeneous in this part of the sample.  Systematic doping via adsorption on patterned graphene could result in a downwards shift of the bands, placing the Fermi level further in the gap.  In order to explore this possibility and gain further insights to the changes of the electronic dispersion we proceed with a more sophisticated analysis where we simulate the $(E,k)$-dependent ARPES intensity and fit the underlying spectral parameters and $E(k)$-dispersion to the measured spectra, as shown in Figure \ref{fig:2}c.  The massless and massive dispersions in pristine and patterned regions, sketched in Figure \ref{fig:2}d,  are modelled using the dispersion relation $E_{\alpha}(k) = s_{\alpha}\sqrt{(\hbar v k)^2 +(m_{\alpha}^{\ast}v^2)^2}$.  Here, $v$ is the band velocity, $m_{\alpha}^{\ast}$ is the effective mass with $\alpha = \{v,c\}$ describing the valence or conduction band and the sign of the dispersion is given by $s_{v}=-1$ and $s_{c}=1$. For $k = 0$ one obtains the band gap $E_g = E_c - E_v = (m_c^{\ast}+m_v^{\ast})v^2$.  For massless quasiparticles in pristine graphene, the expression reduces to the linear dispersion $E(k) = \hbar v k$. The dispersions, which have been overlaid on the spectra in Figures \ref{fig:2}a and  \ref{fig:2}c, are fit by a constant $v = 1.1 \cdot 10^6 \mathrm{~ms}^{-1}$ for both pristine and patterned graphene. The effective mass $m_v^{\ast}$ that defines the curvature of the valence band in patterned graphene is found to increase with hole diameter.

Allowing for a transition from linear to hyperbolic bands between pristine and patterned graphene is required in order to get an optimum $\chi^2$ for the comparison of measured and simulated intensities. The spectra cannot be described by simply allowing for a combination of random doping of the linear pristine graphene dispersion with energy- and momentum-broadenings of the linewidths (for further details see  Figures \ref{fig:S3} and \ref{fig:Doped}).  We can also rule out systematic doping effects in the patterned graphene as the origin of the observed EDC shift $\delta$, as these can not explain the simultaneous increase of the band effective masses.  These changes are instead attributed to the geometry of the lithographic superstructures given by the varying hole diameters $D$.

We are now able to examine the scaling of the induced band structure modifications with $D$. The extracted values of $\delta$ and $m_{v}^{\ast}$ from the independent EDC and dispersion analyses are plotted as a function of hole diameter in Figures \ref{fig:2}e-f.  Both quantities exhibit a linear increase until a breakdown of this trend is reached as the hole diameter approaches the 35 nm period.  At the point where the linear trend breaks down a superstructure effect is still present but it is weaker than expected for the patterned feature sizes. This is likely caused by a combination of a very thin neckwidth of remaining graphene between the holes and growing hole-edge disorder that likely compromises the periodic structure that the 200 nm synchrotron beam measures \cite{Mackenzie2017,Jessen2019}.  Further correlative studies of the microscopic structure around the holes combined with nanoARPES are required to understand the impact of superstructure disorder on the electronic dispersion.  Linear energy and mass scalings with feature sizes have been predicted from antidot lattice models using the Dirac Hamiltonian \cite{Pedersen2008}.  In a simple model using cylindrical geometry to describe the unit cell of the superstructure,  the full energy band gap scales as $E_g = 8\hbar vP^{-2}D$ with $P$ the period,  and equivalently for the band effective masses $m_c^{\ast}+m_v^{\ast} = 8\hbar v^{-1}P^{-2}D$ \cite{Brun2014}.  For $P = 35$ nm one obtains $E_g = 4.3$ meV/nm $\cdot D$ and $m_c^{\ast}+m_v^{\ast} = 0.8 \cdot 10^{-3}$ $m_e$/nm $\cdot D$.  Fits of the extracted values of $\delta$ and $m_v^{\ast}$ to a linear function of $D$ provide an excellent description of the data, excluding the breakdown point,  as seen in Figures \ref{fig:2}e and \ref{fig:2}f. The resulting slopes are $(3.3 \pm 0.1)$ meV/nm and $(0.6 \pm 0.2)\cdot 10^{-3}$ $m_e$/nm for $\delta$ and $m_v^{\ast}$, respectively.  The linear dependence with strikingly similar slopes to those predicted in the simple model for $E_g$ and $m_c^{\ast}+m_v^{\ast}$ strongly suggest that the energy and mass scalings of the induced band structure in patterned regions adhere to antidot lattice scaling relations.  Note that we cannot experimentally access the exact dependence on hole diameter as we only measure the part of the gap that is shifted below the Fermi energy and we are not able to determine $m_c^{\ast}$ as the conduction band is unoccupied at this doping level.  

We can rule out gap-opening effects stemming from disorder-mediated charge carrier localization \cite{Stampfer:2009} by extracting the quasiparticle mean free path for pristine and patterned graphene regions using the linewidth of momentum distribution curves (MDCs) of the corresponding ARPES spectra, as shown in Figure \ref{fig:S5}.  The extracted MDC linewidths at a binding energy of 0.45 eV range from 0.075 to 0.096~\AA$^{-1}$ for pristine and patterned graphene regions, which are similar to linewidths reported by ARPES on graphene/hBN devices \cite{Muzzio2020} and reasonable when compared to pristine high-quality graphene synthesized on silicon carbide \cite{Bostwick2007}.  The mean free path is shorter in the patterned regions,  indicating an overall higher level of scattering than in pristine graphene.  However,  the change in scattering between different patterned graphene regions does not correlate with the increasing effective mass and shift of the EDC leading edge and can therefore not contribute to the observed spectral changes.

\begin{figure*} [t!]
	\begin{center}
		\includegraphics[width=1\textwidth]{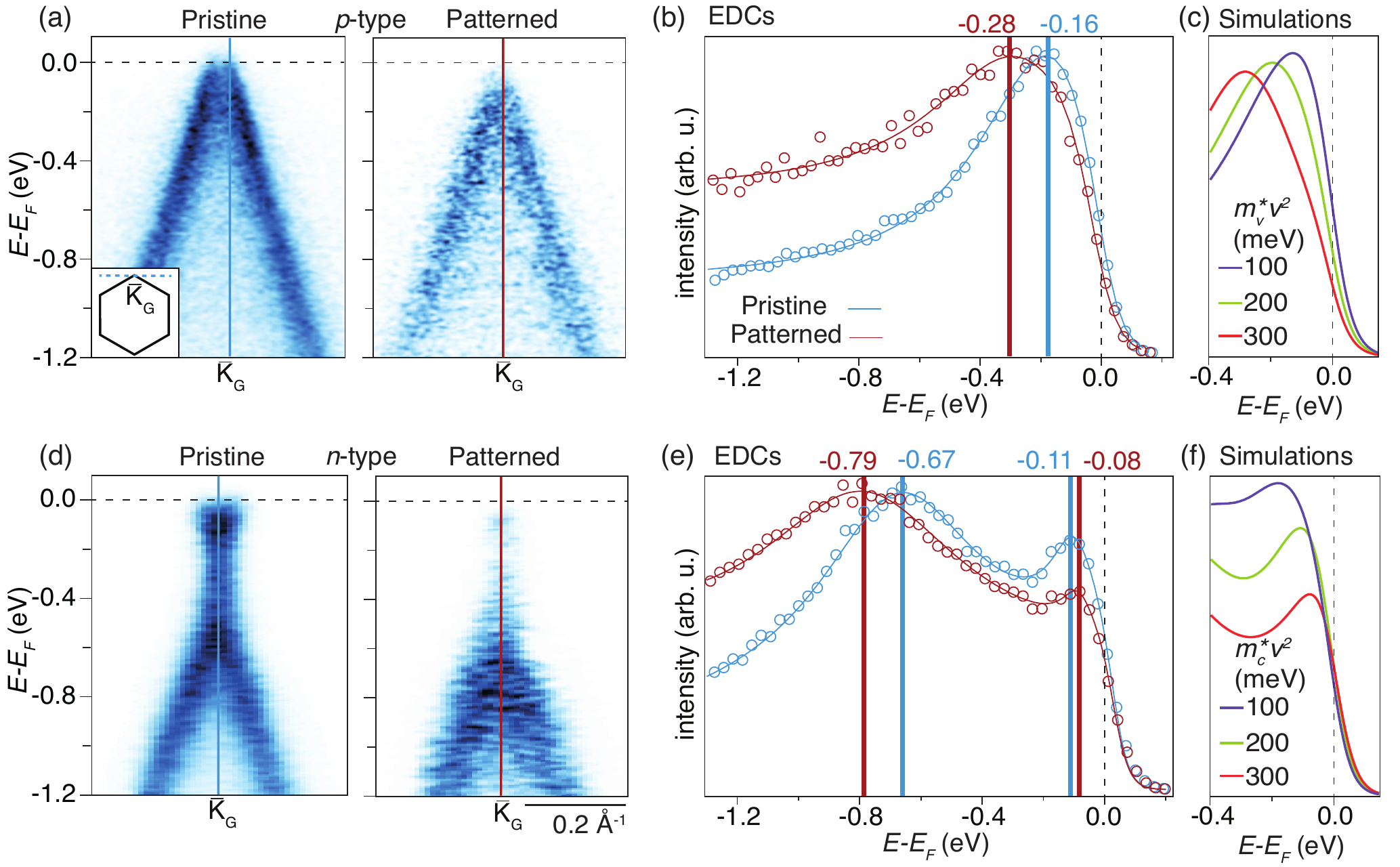}
		\caption{Electrical tunability of patterned graphene.  a) ARPES intensity of pristine and patterned graphene for $p$-type doping of $(0.7 \pm 0.2)\cdot 10^{12}$~cm$^{-2}$.  b) EDCs (markers) extracted from (a) at $\bar{\mathrm{K}}_\mathrm{G}$ for patterned graphene and at the Fermi level crossing of the linear branch for pristine graphene.  Fits (curves) to a Lorentzian peak multiplied by a Fermi-Dirac distribution on a linear background provide the noted peak positions marked by vertical bars.  c) Simulated EDCs at $\bar{\mathrm{K}}_\mathrm{G}$ for a hyperbolic valence band,  illustrating the evolution of EDC energy-dependence and peak position with increasing mass term.  d)-f) Corresponding data for $n$-type doping of $(4.0 \pm 0.6)\cdot 10^{12}$~cm$^{-2}$.  Both EDCs in (e) have been extracted at $\bar{\mathrm{K}}_\mathrm{G}$ and an additional Lorentzian peak is included in the fit to account for the upper branch.  The simulated EDCs in (f) include the hyperbolic conduction band with the stated mass terms.  The ARPES spectra have been acquired along the cut shown in the BZ inset in (a).  The error bars on the stated peak positions are $\pm 0.03$ eV,  corresponding to one standard deviation of the Lorentzian peak fits.}
		\label{fig:3}
	\end{center}
\end{figure*}

In order to estimate the magnitude of $E_g$ we shift the bands determined in the ungated $p$-type situation in \textbf{Figure \ref{fig:3}}a to lower energies by utilizing the back-gate in our device architecture to induce electron-type ($n$) doping. The dispersion for the maximum achievable $n$-doping before breakdown of the device is presented in Figure \ref{fig:3}d. The Dirac point region in pristine graphene becomes visible and linear extrapolation of the bands leads to $k_F = (0.035 \pm 0.003)$~\AA$^{-1}$ at $E_F$, corresponding to an $n$-type carrier concentration of $(4.0 \pm 0.6) \cdot 10^{12}$~cm$^{-2}$.  By comparison, the dispersion of $n$-type patterned graphene is characterized by a hyperbolic valence band with a band maximum shifted down in energy and an elevated level of intensity around the Fermi energy that is consistent with occupied conduction band states.  

EDCs extracted around $\bar{\mathrm{K}}_\mathrm{G}$ provide quantitative information on the magnitude of the band shifts and  dispersion changes between pristine and patterned regions in the two doping scenarios.  At the $p$-type doping in Figure \ref{fig:3}b we observe a shift of the leading edge,  as discussed in connection with Figure \ref{fig:2}b, as well as a peak shift of $(0.12 \pm 0.03)$ eV to lower energies between the EDCs of pristine and patterned regions.  Furthermore, the energy-dependence of the EDC tails to lower energies are distinctly different in the two situations.  These spectral changes are examined by extracting EDCs from simulations of the ARPES intensity emerging from a hyperbolic dispersion given by $m_v^{\ast}v^2$-terms in the range of 100-300 meV as shown in Figure \ref{fig:3}c.  The increasing mass shifts the hyperbolic band to lower energies and flattens it, leading to the observed peak shifts and intensity tails.  

In the $n$-doped situation the peak shift in the valence band region persists while the conduction band peak in the patterned graphene is situated closer to the Fermi energy than in the pristine graphene,  as seen in Figure \ref{fig:3}e.  This behavior is explained by simulated EDCs of the hyperbolic dispersion in the conduction band region in Figure \ref{fig:3}f.  A larger $m_c^{\ast}v^2$-term shifts the conduction band peak closer to the Fermi energy.  Larger masses of the valence and conduction bands lead to increasing separation of the corresponding EDC peaks, thereby widening the band gap of the system.  In Figure \ref{fig:3}e the energy difference between the EDC peak positions increases for the patterned graphene compared to pristine graphene,  consistent with the opening of a band gap.  There is no shift of the EDC leading edge in the $n$-doped situation because the occupied conduction band states of the patterned graphene cross the Fermi energy. 

A diffuse tail of intensity is visible between the band extrema in patterned graphene,  similarly as observed in previous gap measurements of graphene/hBN superlattices using ARPES \cite{Wang:2016}. The residual intensity is likely caused by inhomogeneously broadened defect states.  It has been shown that lattice vacancies and bond reconstructions contribute additional spectral weight localized in $k$ around $\bar{\mathrm{K}}_\mathrm{G}$ \cite{Kot2020}.  Similar features have been reported for pattern-hydrogenated graphene and attributed to an imbalance between the graphene sublattices \cite{Grassi:2011}.  Our 200 nm beam averages inhomogeneously over such defects, which are anticipated around edges of the patterned holes  \cite{Mackenzie2017,Jessen2019,Power2014}.  It is important to note that defective graphene is characterized by a stretched Dirac point with an elevated level of photoemission intensity throughout the stretched part of the spectrum \cite{Rotenberg:2008,Kot2020}. In our case,  the significant loss of intensity between the band extrema and the shift of the EDC leading edge observed in Figure \ref{fig:2}b can only be explained by the opening of a band gap.

\begin{figure*} [t!]
	\begin{center}
		\includegraphics[width=0.5\textwidth]{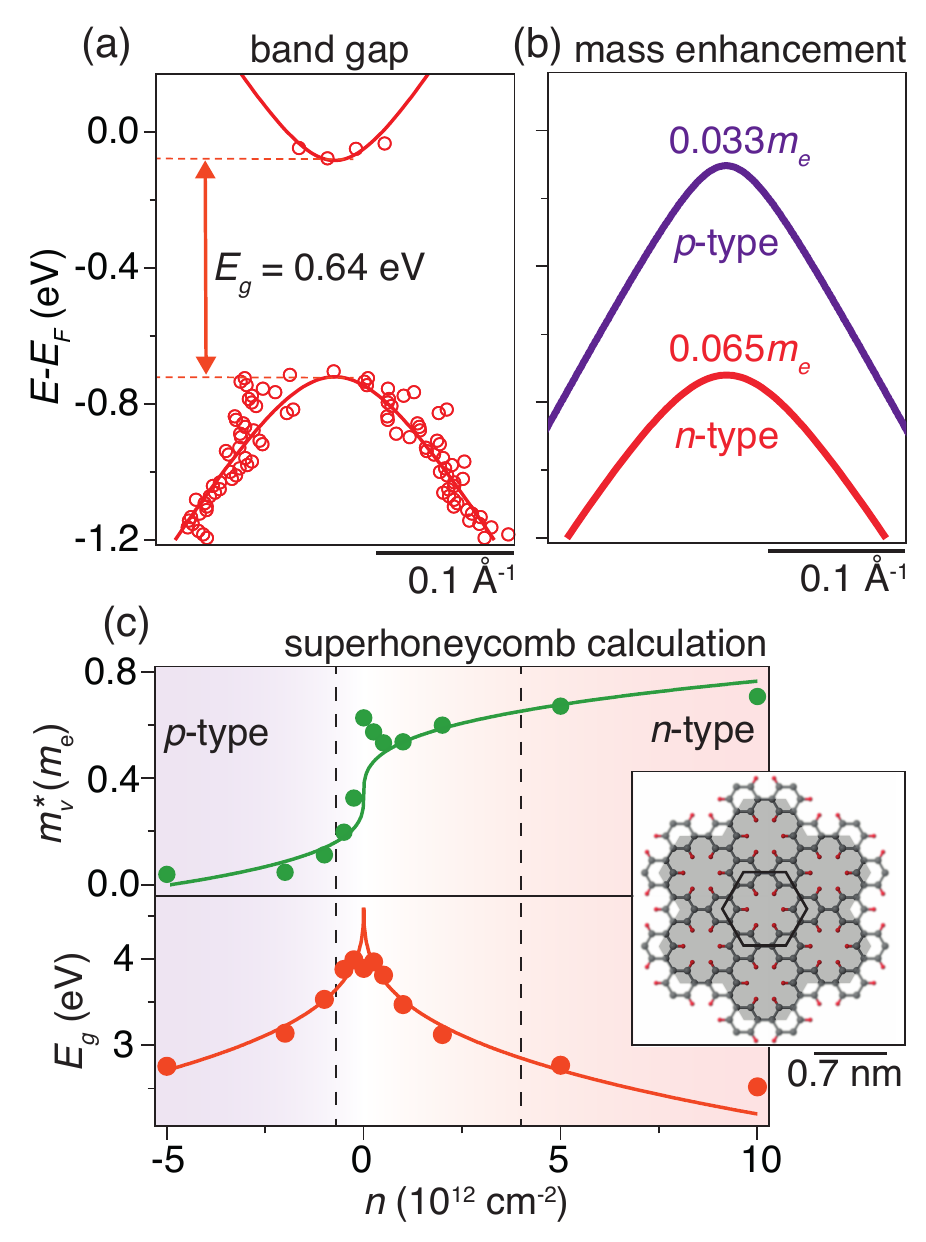}
		\caption{Doping-dependence of band parameters.  a) Peak positions (markers) from combined EDC and MDC analysis of the ARPES intensity of patterned graphene in Figure \ref{fig:3}d with fits (curves) to hyperbolic valence and conduction bands. The band gap, $E_g$, is marked by a double-headed arrow. The error bar on the gap is $\pm 0.03$ eV,  corresponding to one standard deviation of the peak position fits.  b) Extracted valence band dispersion and effective masses for $p$- and $n$-type dopings.  The error bars on the effective masses derived from the dispersion fits are $\pm 0.003$ $m_e$. c) Doping-dependence of effective hole mass and band gap for an approximate antidot lattice structure (see inset) determined from $\textrm{G}_0\textrm{W}_0$ calculations (markers). Curves represent fits to a quasi-relativistic two-band model. Vertical dashed lines mark the doping levels in the nanoARPES experiment.}
		\label{fig:4}
	\end{center}
\end{figure*}

In order to achieve a more precise estimate of the peak-to-peak energy of the induced band gap in $n$-doped patterned graphene we extract the dispersion from combined EDC and MDC fits.  By fitting the combined peak positions to hyperbolic valence and conduction bands we obtain a band gap of $E_g = (0.64 \pm 0.03$)~eV, as shown in \textbf{Figure \ref{fig:4}}a  A wide energy-tunability of the gap is demonstrated by the $(0.55 \pm 0.03$)~eV shift of the valence band in addition to a doubling of effective hole mass between the two doping levels, as seen in Figure \ref{fig:4}b.  It is interesting to compare the magnitude of the observed band gap with previous transport measurements that reported a gap of ($148 \pm 22$) meV on a patterned graphene device with similar feature sizes \cite{Jessen2019}.  This transport gap was measured in a device encapsulated in hBN, whereas our sample is exposed to vacuum on one side.  The gap measured by transport is therefore more strongly renormalized due to environmental screening.  Furthermore, the transport gap is determined by varying the doping and sweeping the valence and conduction band edges through the Fermi energy, whereas we measure at fixed $n$-doping where many-body effects may further renormalize the band parameters.  Finally, it is not well-established how the peak-to-peak energies reported from the ARPES measurement translate to the sensitivity of the transport measurement and the modelling applied to estimate a gap.  Further gap measurements of nano-engineered samples using nanoARPES in direct comparison with transport measurements on the same samples are an important future avenue to explore in order to establish a clear link between the two approaches.

The large renormalisation of hole masses in antidot superlattices is a result of the pronounced impact of doping on screening in two-dimensional semiconductors. Similarly large effects are expected for the band gap, which we are only able to observe in ARPES at high $n$-doping. To substantiate the expected doping-dependence of effective masses and estimate the associated band gap renormalisation, we therefore perform quasiparticle calculations for doped polyphenylene superhoneycomb structures \cite{Aijun:2010}, see inset of Figure \ref{fig:4}c.  Such “porous graphene” can be produced using self-assembly of precursor molecules \cite{Bieri:2009} and may be viewed as the smallest graphene antidot lattice prototype. Due to the reduced size, first-principles quasiparticle calculations are feasible for this structure. However, band gaps and effective masses are necessarily larger than those of the much greater lithographically defined structures. We note that carrier-induced band structure renormalisation cannot be described at the density-function theory (DFT) level, as it is intimately connected to quasiparticle corrections. We therefore performed G$_0$W$_0$ calculations on top of DFT band structures (see Figure \ref{fig:S6}) in order to extract band gaps and effective masses, as shown in Figure \ref{fig:4}c. The trends outside the weak-doping regime can be captured by a quasi-relativistic two-band model, described in Supplementary Note 8,  suggesting a symmetric $|n|^{1/3}$ -dependence of the band gap, whereas the hole mass varies as sign($n$)$|n|^{1/3}$.  As a consequence, a step-like increase of the hole mass appears as doping shifts from $p$ to $n$-type, in line with the experimentally observed doubling of the mass in Figure \ref{fig:4}b. Importantly, the calculations clearly imply a larger gap at low carrier concentrations compared to the one observed in the high-density regime probed by nanoARPES. In the superhoneycomb structure, an increase of $\sim30\%$ is observed as $n$-doping is reduced from $4.0 \cdot 10^{12}$~cm$^{-2}$ to zero. This suggests that the already substantial observed gap of ($0.64 \pm 0.03$)~eV is a result of renormalisation of an even wider intrinsic gap, which is tunable both via carrier doping and geometry of the lithographically-defined superlattice.

\section{Conclusion}
Our work demonstrates the manner in which lithographically engineered superstructures dramatically impact the electronic structure and lead to tunable massive Dirac fermions in a patterned graphene device.  By directly measuring the hyperbolic quasiparticle dispersion and its electrical tunability with energy,  momentum,  and spatial resolution using nanoARPES, we demonstrate a versatile platform for tailoring and characterizing electronic structures at the nanoscale.  We observe a substantial quasiparticle band gap of ($0.64 \pm 0.03$)~eV, which introduces important new questions pertaining to how the gap size is affected by many-body interactions, including impurities and structural defects that inevitably form during patterning, as well as its utility in electron transport and optical excitations.  Further correlative microscopic and spectroscopic investigations combined with quasiparticle calculations on the same superstructures are therefore required to uncover the full potential of band structure engineering with lithographic structuring of 2D materials.  Our observations of linear energy and mass scalings with feature sizes underline the Dirac nature of the engineered superstructures in graphene, which open a vast number of opportunities for deterministic, local control of electronic properties with the potential to overcome the limitations associated with lattice matching conditions, twist angle, and scalability in common superlattice designs.

\section{Experimental Section}

\textit{Heterostructure assembly.} Mechanically exfoliated flakes of graphene and hBN (44\,nm) were transferred on top of a graphite (14\,nm) back-gate using cellulose acetate butyrate (CAB) \cite{Schneider:2010,Thomsen:2017}. The graphite was exfoliated onto 90\,nm SiO$_2$ on intrinsic silicon to minimize the risk of dielectric breakdown. The CAB was removed by soaking the samples in ethyl acetate for one hour, then 30 minutes in acetone, and finally rinsed in IPA. Subsequently, the samples were annealed in forming gas at 350\,$^{\circ}$C for 3 hours. \\

\textit{Lithography.} All electron beam lithography (EBL) was carried out in a 100\,keV JEOL JBX-9500FS system. For fabricating contacts to the graphene and graphite, PMMA (996k, 5\% by weight in anisole) was spin-coated at 2000 rpm for one minute, and subsequently baked for 10 minutes at 180\,$^{\circ}$C. The PMMA was exposed using a dose of $900\,\mu$C/cm$^2$ and developed using IPA:H$_2$O (3:1) under ambient conditions for 60 seconds. Subsequently Cr/Au (5/100$\,$nm) contacts were e-beam deposited at a pressure of $<$4$\cdot 10^{-8}$\,Torr, and lift-off was done in acetone. To ensure good electrical contact, several electrodes were added to both the graphene and the graphite gate. Images of the stacking sequence and the final device are presented in Figure \ref{fig:DevFig}.

For superstructure fabrication, the devices were baked for 10 minutes at 180\,$^{\circ}$C, followed by spin-coating of PMMA (2200k, 1\% by weight in anisole), and subsequently baked for one hour at 180\,$^{\circ}$C, resulting in a 47\,nm thick resist. The periodic patterns of holes were exposed using doses of 1-6\,fC/hole and a beam-current of 0.8\,nA at 100$\,$kV. Development was done using pure IPA for 30 seconds, followed by nitrogen blow-drying. Finally, the samples were etched with reactive ion etching for 15 seconds, using a gas of O$_2$ at 80\,mTorr and a forward power of 20\,W. On a separate device we measured 10.2\,nm of PMMA being etched using this process. Except for the lack of a thin top hBN layer, the above process mirrors that described in Ref. \cite{Jessen2019}.\\

\textit{NanoARPES experimental details.} The measurements were performed at the Microscopic and Electronic Structure Observatory (MAESTRO) at the Advanced Light Source, Lawrence Berkeley National Laboratory, USA. Prior to measurements, samples were annealed at 120 $^{\circ}$C for 6 hours in ultra-high vacuum (UHV) in order to desorb water from the transfer in air. The nanoARPES measurements were performed at a base pressure better than  $1 \cdot 10^{-10}$~Torr and with the sample held at room temperature.

The synchrotron beam-spot was focused to 200 nm at a photon energy of 97 eV using Fresnel zone plate optics in combination with an order sorting aperture. Alignment of optics for focusing and precision mapping of the photoemission intensity was achieved using a coarse piezo-scanning stage in combination with a flexural piezo-scanner for finer stepping. The smallest scan step size was 100 nm, oversampling our spatial resolution for photoemission by a factor two. The data were collected using a hemispherical Scienta R4000 electron analyzer with energy- and angular-resolution better than 30 meV and 0.1$^{\circ}$, respectively. In order to obtain an $E(k)$-cut that passes exactly through $\bar{\mathrm{K}}_{\mathrm{G}}$, first the rough position of $\bar{\mathrm{K}}_{\mathrm{G}}$ was determined by rotating the detector around the sample in 0.1\degrees~steps over a range of $\approx$2\degrees~with a low acquisition time. From this rough position, a fine measurement with an acquisition time of 40-60 minutes was made by rotating the detector in steps of 0.1\degrees~over a small range of 0.3\degrees-0.5\degrees~around $\bar{\mathrm{K}}_{\mathrm{G}}$, enabling a correct $k$-transformation and extracting the $\bar{\mathrm{K}}_{\mathrm{G}}$-cut. Note that the sample position is held fixed relative to the synchrotron beam for such angle scans, such that we are able to measure the same area without any drift.

For electrostatically gated measurements we applied a voltage between the graphite contact and the top contact on the graphene until a clear $n$-doping effect in ARPES measurements of the Dirac cone of pristine graphene was observed. In order to correct for the electric field displacement of photoelectrons with a finite gate, we performed  angle scans following the procedure described above in order to locate and precisely measure the intensity along the $\bar{\mathrm{K}}_{\mathrm{G}}$-cut. Prolonged application of a gate voltage combined with exposure to the synchrotron beam led to an irreversible increase of the background intensity, limiting gated measurements on multiple patterned regions.\\

\textit{Calculations.} The $\textrm{G}_0\textrm{W}_0$ and underlying density functional theory (DFT) band structures were both calculated using the GPAW library \cite{Mortensen:2005,Enkovaara:2010,Huser:2013}. DFT band structure calculations based on relaxed atomic positions and unit cells were performed using the PBE exchange and correlation functional \cite{Perdew:1996} on an 18$\times$18 $k$-point $\mathrm{\Gamma}$-centered Monkhorst-pack grid, using 350 DFT bands, a plane wave basis with a 500 eV cutoff, 1 meV Fermi occupation smearing extrapolated to 0 K, and otherwise default GPAW parameters, ensuring adequate convergence. Finally, a vacuum slab of 10 Å was added in the out-of-plane direction, ensuring an isolated 2D structure. Six $\textrm{G}_0\textrm{W}_0$ bands, three in the valence and three in the conduction band, were subsequently computed on top of the DFT calculations, utilizing an analytical correction to the $q=0$ term and a 2D Coulomb truncation to avoid long-range interactions between periodically repeated unit cells in the out-of-plane direction \cite{Huser:2013}. The $\textrm{G}_0\textrm{W}_0$ self-energy was extrapolated to an infinite plane wave cutoff energy based on three lower values (20, 30 and 40 eV) \cite{Huser:2013}. These rather low cutoff energies are imposed by computational demands but deemed sufficient since low energy plane waves dominate any band structure renormalization due to doping \cite{Gao:2017}. The $n$- and $p$-doping densities were introduced by requiring charge neutrality after including a 7~Å thick periodic jellium slab of the opposite charge. The effective masses were obtained from a quasi-relativistic fit inside a small circular region containing 31 $k$-points around $\bar{\mathrm{K}}$.\\

\textit{Statistical Analysis.} The raw nanoARPES spectra were acquired as a function of electron kinetic energy and emission angle.  The energy scale was calibrated to the Fermi edge on the gold contact and all datasets were transformed from angle to momentum coordinates using Igor Pro 7 from WaveMetrics.  Simulations of ARPES intensity,  data analysis and fits of line profiles were performed using the same software.  The fit in Figure \ref{fig:1}e was performed using a square wave convoluted by the spatial resolution transfer function, which was taken as a Gaussian with a full-width half maximum of 200 nm.  The mean values of $I_{min}/I_{max}$ resulting from the fit are displayed in Figure \ref{fig:1}g with error bars representing one standard deviation. The error bars on $D$ have been propagated from $I_{min}/I_{max}$. The fits in Figure \ref{fig:2}b were composed of a Fermi-Dirac function multiplied by a linear background.  Resulting mean values of the shift of the mid-point of the fitted distributions are shown in Figure \ref{fig:2}e with error bars representing one standard deviation, as obtained from the fits. The effective masses in Figure \ref{fig:2}f were obtained by comparing the simulated photoemission intensity to the measurements and minimizing $\chi^2$ as described in Supplementary Note 5 in the Supporting Information. The error bars represent one standard deviation determined from the $\chi^2$-distribution.  EDC fits in Figures \ref{fig:3}b and \ref{fig:3}e were performed using Lorentzian peaks multiplied by a Fermi-Dirac function on a linear background.  Mean values of fitted Lorentzian peak positions are reported with an error bar of $\pm 0.03$ eV corresponding to one standard deviation.

\section{Acknowledgement}
S. U. acknowledges financial support from the Independent Research Fund Denmark under the Sapere Aude program (Grant No. 9064-00057B) and VILLUM FONDEN under the Young Investigator Program (Grant No. 15375). This research used resources of the Advanced Light Source, a U.S. DOE Office of Science User Facility under contract no. DE-AC02-05CH11231. P. B. , A. -P. J. , L. G., M.O.S. and T.G.P.  acknowledge support from Danish National Research Foundation (DNRF) Center for Nanostructured Graphene (Grant  No. DNRF103), EU Graphene Flagship Core 2 (Grant No. 785219) and Core 3 (Grant No. 881603). Growth of hexagonal boron nitride crystals was supported by the Elemental Strategy Initiative conducted by the MEXT, Japan, Grant Number JPMXP0112101001 and  JSPS KAKENHI Grant Number JP20H00354. Nano-lithography and nano-ARPES work at Columbia are supported as part of Programmable Quantum Materials, an Energy Frontier Research Center funded by the U.S. Department of Energy (DOE), Office of Science, Basic Energy Sciences (BES), under award DE-SC0019443. The authors acknowledge Dmitri Basov for helpful comments on the manuscript.

\newpage

\noindent\textbf{Supplementary Note 1.  Composition of nanostructured device}

\begin{figure*} [h!]
	\begin{center}
		\includegraphics[width=1\textwidth]{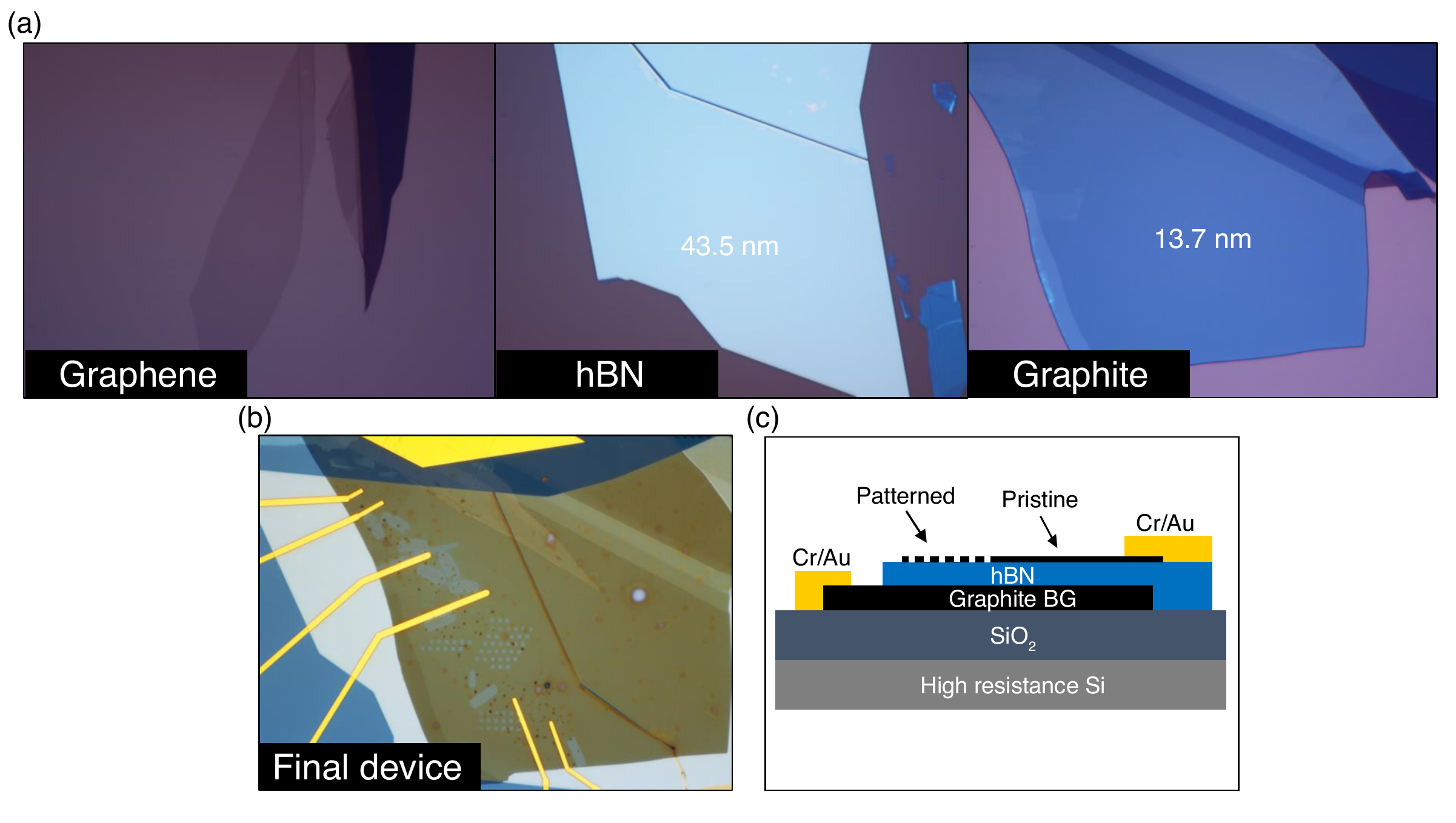}
		\vspace*{-0.8cm}
		\caption{Overview of device components.  a) Optical images of graphene, hBN and graphite. The thicknesses of hBN and graphite flakes are stated.  b) Optical micrograph of the device.  c) Diagram of the vertical device structure with key components labelled.}
		\label{fig:DevFig}
		\vspace*{-0.8cm}
	\end{center}
\end{figure*}

The nanopatterned graphene device was built from mechanically exfoliated graphene and hBN layered on top of a graphite back gate. These individual components are shown in Figure \ref{fig:DevFig}a prior to stacking. After stacking, a variety of superstructures were patterned into the graphene, followed by evaporation of electrical contacts to allow for gating and grounding of the sample. The optical image in Figure \ref{fig:DevFig}b shows the device. The vertical structure of the device is laid out diagrammatically in Figure \ref{fig:DevFig}c. The graphite back gate and graphene are separately contacted to allow for electrical doping. 
 
\newpage

\noindent\textbf{Supplementary Note 2.  Twist angle between graphene and hBN}

\begin{figure*} [b!]
	\begin{center}
		\includegraphics[width=0.8\textwidth]{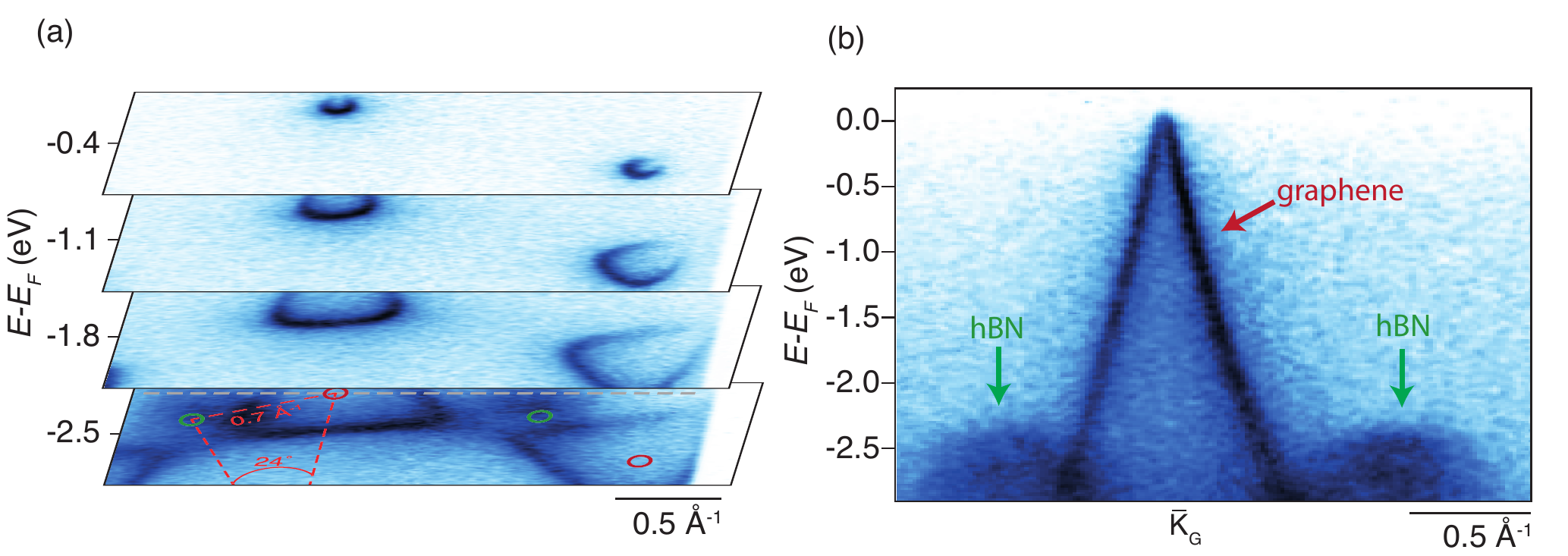}
		\caption{ARPES dispersion of graphene and hBN.  a) Constant energy slices extracted at the labelled energies. Two graphene (hBN) $\bar{\mathrm{K}}$-points have been marked with red (green) circles in the measured segment of the BZ. The red dashed lines indicate the geometry used to determine the twist angle between hBN and graphene.  b) $(E,k)$-dependent ARPES intensity obtained along the dashed grey line shown in the bottom panel of (a), highlighting graphene and hBN bands.}
		\label{fig:S1}
	\end{center}
\end{figure*}

To determine the Brillouin zone (BZ) alignment and twist angle between the graphene and hBN, ARPES measurements were performed with an achromatic capillary optic with a beam-spot size of $\approx$1.5 $\mu$m \cite{Koch2018}. The high transmission of this optic enables rapid pre-alignment of the sample prior to the low-flux zone plate measurements with a $\approx$200 nm beam. Figure \ref{fig:S1} presents photoemission spectra obtained from a pristine region of the sample. The data were measured by rotating the detector around the sample and capturing a photoemission spectrum at each polar emission angle, such that the $(E,k_x,k_y)$-dependent intensity could be obtained. 

Figure \ref{fig:S1}a presents constant energy slices extracted from such a dataset. At higher energies, the characteristic horseshoe-shaped constant energy contours of the graphene ARPES dispersion are visible \cite{Bostwick2007}. At -2.5 eV, additional pockets of intensity become apparent, which arise from the valence band maximum of the underlying hBN. The dispersion of these features are presented in the example $E(k)$-cut around a $\bar{\mathrm{K}}_{\mathrm{G}}$-point in Figure \ref{fig:S1}b. The interlayer twist angle is determined using $\theta=\arccos\left[(G_{\mathrm{G}}^2+G_{\mathrm{BN}}^2-G_{\mathrm{\Delta}}^2)/2G_{\mathrm{G}}G_{\mathrm{BN}}\right]$, where $G_{\mathrm{G}}$ ($G_{\mathrm{BN}}$) is the reciprocal lattice vector of graphene (hBN) and $G_{\Delta}$ is the $k$-separation of the $\bar{\mathrm{K}}_{\mathrm{G}}$- and $\bar{\mathrm{K}}_{\mathrm{BN}}$-points demarcated by red and green circles in Figure \ref{fig:S1}a, respectively. We determine $G_{\Delta} = (0.7 \pm 0.04)$ Å$^{-1}$, leading to a twist angle of ($24 \pm 0.5$)\degrees.

\newpage

\noindent\textbf{Supplementary Note 3.  Spatially-dependent photoemission intensity}

\begin{figure*} [b!]
	\begin{center}
		\includegraphics[width=1\textwidth]{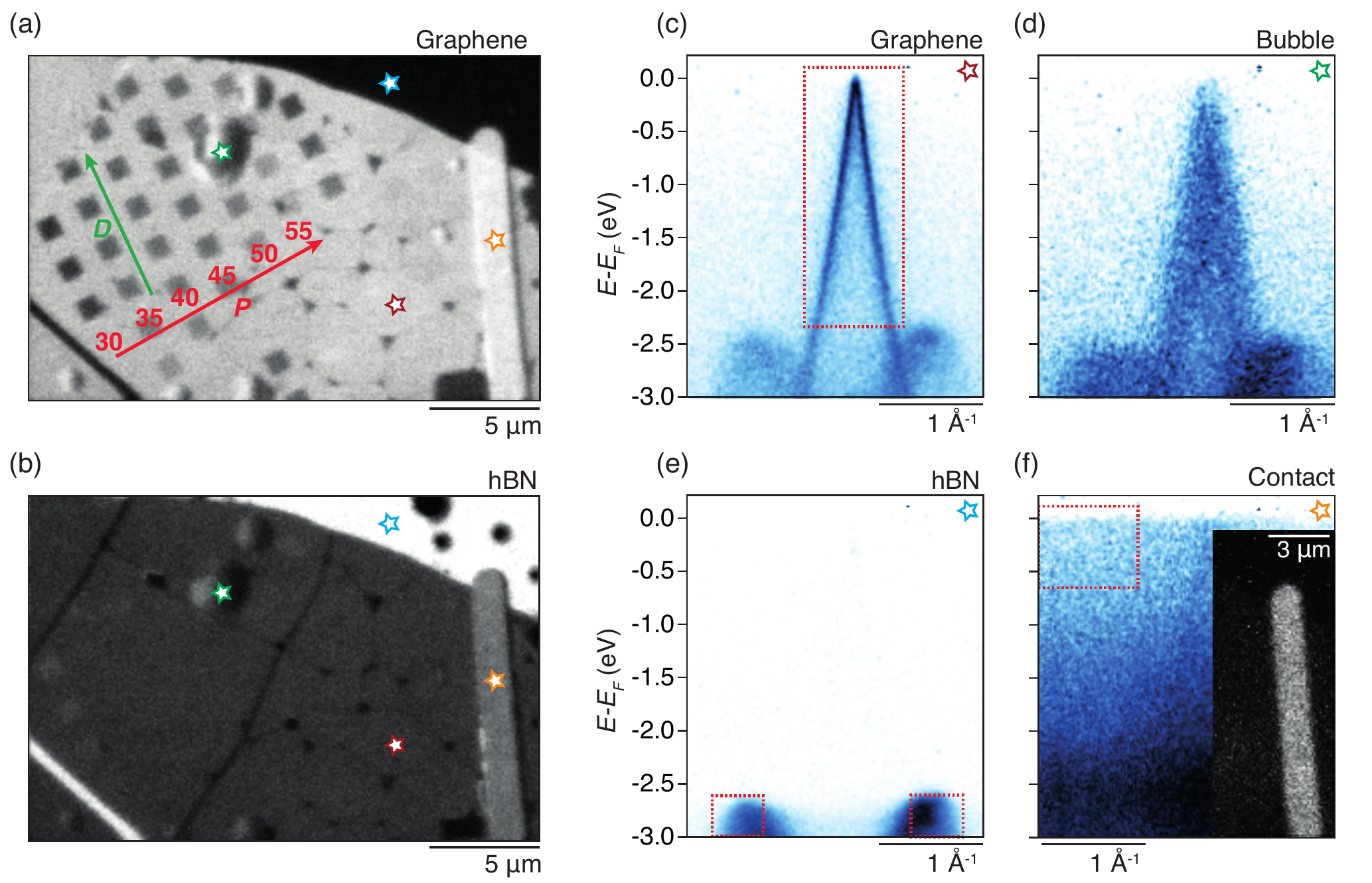}
		\vspace*{-0.8cm}
		\caption{Spatially varying photoemission intensity.  a) Dirac cone $(E,k)$-intensity projected onto the $(x,y)$-coordinates of an area with patterned and pristine graphene. The step-size of the scan is 100 nm. The periodicity $P$ of holes varies along the columns indicated by a red arrow, and the value of $P$ is given in units of nanometers. The hole diameter increases in the direction of the green arrow.  b) Map of the same region composed from the intensity of the hBN valence band.  c)-f) ARPES snapshots from the areas marked by correspondingly coloured stars in (a)-(b). Dashed boxes in (c) and (e) demarcate the regions integrated to produce (a) and (b) respectively.  f) ARPES intensity of the electrical contact. The area within the dashed red box is integrated to produce the map of the contact in the inset.}
		\label{fig:S2}
		 		\vspace*{-0.4cm}
	\end{center}
\end{figure*}

Here we provide additional details of the four-dimensional nanoARPES intensity map shown in Figure 1d of the main manuscript. Figure \ref{fig:S2}a shows the spatial distribution of spectral intensity integrated around the Dirac cone of graphene. As described in connection with Figure 1 of the main text, the dark squares are regions that have been lithographically patterned. Along each column of squares the periodicity of holes, $P$, varies according to the labelled values stated in units of nanometer, while along each row the hole diameter increases. Numerous additional features stand out in the nanoARPES intensity map, including a circular area, marked with a green star, and a dark patch in the top of the map, marked in blue. The origin of these features is discussed below by investigating the $(E,k)$-dependent photoemission intensity arising from these microscopic areas. The spatial distribution of Dirac cone intensity is contrasted to Figure \ref{fig:S2}b where the intensity integrated around the valence band maximum of hBN has been mapped. Regions that are dark in Figure \ref{fig:S2}a now exhibit a high intensity. Other features, such as the web of dark lines around the red star are present in both maps.  These are attributed to wrinkles that percolate the flakes.

Figures \ref{fig:S2}c-f present the $E(k)$-dispersion extracted from 1 $\times$ 1 $\mu$m$^2$ areas marked by correspondingly coloured stars in Figures \ref{fig:S2}a-b. In the region marked by a red star, the linear graphene bands are observed in addition to a faint valence band intensity from the underlying hBN. The area marked by a green star displays a dramatically broadened Dirac cone. This is attributed to a bubble in the graphene, which gives rise to a continually varying normal emission angle across its surface. Our integration region thereby sums intensity from many different angles, causing such a broadening. The top area of the probed region, marked by a blue star, exhibits intense valence bands of hBN \cite{Soren2018_hBN} due to the absence of the overlying graphene flake in this region. Figure \ref{fig:S2}f displays a homogeneous intensity distribution with a flat Fermi edge. Integrating the intensity around the Fermi level and projecting onto the real-space coordinates of the scan reveals the map in the inset of Figure \ref{fig:S2}f, corresponding to a polycrystalline contact in this area of the device.\\

\newpage

\noindent\textbf{Supplementary Note 4.  Details of energy distribution curve analysis}

\begin{figure*} [b!]
 	\begin{center}
 		\includegraphics[width=1\textwidth]{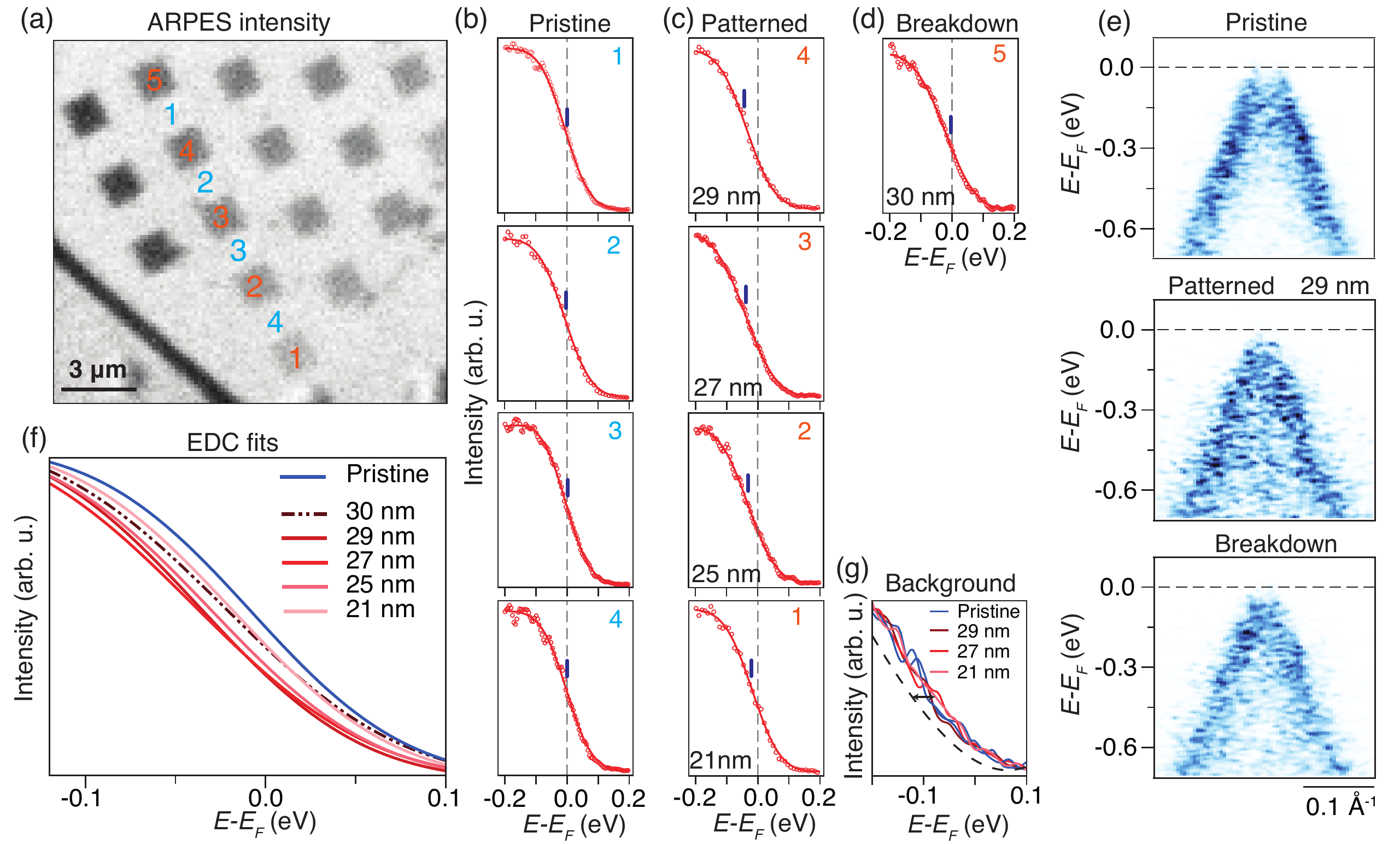}
 		\caption{Details of EDC analysis to extract shifts of the band edge in patterned graphene.  a) Close-up of the photoemission intensity map from Figure \ref{fig:S2}a.  Blue (red) numbers label areas selected for measurements of pristine (patterned) graphene.  b)-d) EDCs from the noted positions integrated $\pm$0.05 $\textrm{Å}^{-1}$ with respect to $\bar{\mathrm{K}}_{\mathrm{G}}$ (markers) with fits (curves) to a Fermi-Dirac function multiplied by a linear energy dependence. Ticks denote the fitted mid-point of the band edge. The Fermi level is marked with a dashed line.  e) Example ARPES spectra from the noted regions.  f) Fits of the leading edge from EDCs on patterned and pristine graphene combined to highlight the shift in leading edge with hole diameter.  g) EDCs from the labelled regions taken at $k$-values where only background states are present.  A polynomial fit (black dashed line) to the background intensity is offset by 45 meV. }
 		\label{fig:S4}
 	\end{center}
 \end{figure*}
 
In order to ensure that the shift in the leading edge of the bands between pristine and patterned graphene, observed in Figure 2b of the main paper, is due to the increasing diameter of patterned holes and not caused by doping or external shifts of the Fermi level, multiple pristine graphene regions were measured adjacent to the areas of patterned graphene.  Figure \ref{fig:S4}a shows a zoom-in on the measured region of the sample, with blue (red) numbers labelling the pristine (patterned) graphene regions measured. Energy distribution curves (EDCs) integrated $\pm$0.05 $\textrm{Å}^{-1}$ with respect to $\bar{\mathrm{K}}_{\mathrm{G}}$ from these regions are shown in Figures \ref{fig:S4}b-d, and example ARPES spectra are presented in Figure \ref{fig:S4}e.  Each EDC is fitted to a Fermi-Dirac function multiplied by a linear energy term. Fits of EDCs from pristine graphene confirm that the Fermi edge is fixed across the studied region of the sample. The EDCs for the patterned graphene are fitted in an identical manner and demonstrate a systematic shift of the leading band edge to lower energies with increasing hole diameter in Figure \ref{fig:S4}c, until the point of breakdown of the patterns in Figure \ref{fig:S4}d. All fits are summarized in Figure \ref{fig:S4}f to highlight the extracted shifts, which are plotted against hole diameter in Figure 2e of the main paper.  

To ensure that these shifts do not arise from spatially dependent charging effects,  EDCs taken at $k$ where only background states are present are inspected for different patterned and pristine regions as shown in Figure \ref{fig:S4}g.  No shifts are observable in the background between these various regions of the sample, indicating that photoelectrons emitted from these regions see the same electric field.  For comparison, a polynomial fit to an EDC of the background intensity is offset by 45 meV, the largest shift determined for the graphene states in the patterned region with $D = 29$ nm, demonstrating how an energy shift of this magnitude would be clearly visible in the background.

\newpage

\noindent\textbf{Supplementary Note 5.  Simulations of photoemission intensity}

In order to extract the valence band dispersion and effective mass, simulations of the photoemission intensity were optimised to the spectra acquired around $\bar{\mathrm{K}}_\mathrm{G}$. The ARPES intensity is given by $I(E,k) = MA(E,k)f(E)$ \cite{Bostwick2007}, where $M$ is taken as a constant and $f(E)=1/(\exp[(E-\mu)/k_BT]+1)$ is the Fermi-Dirac distribution describing the occupation of the bands for a given temperature $T$ and doping $\mu$. The spectral function is written as

\begin{equation}
	A(E,k)=\frac{\pi^{-1}\Sigma''}{[E-E_v(k)]^2+(\Sigma'')^2},
\end{equation}
where $\Sigma''$ is the imaginary part of the self energy, which is approximated by a constant, and $E_v(k) = -\sqrt{(\hbar v k)^2+(m_{v}^\ast v^2)^2}$ is the hyperbolic dispersion of the valence band with velocity $v$ and effective mass $m_{v}^\ast$. The simulated intensity is convoluted with Gaussians to account for experimental energy- and momentum-broadening. 
 
\begin{figure*} [h!]
 	\begin{center}
 		\includegraphics[width=0.85\textwidth]{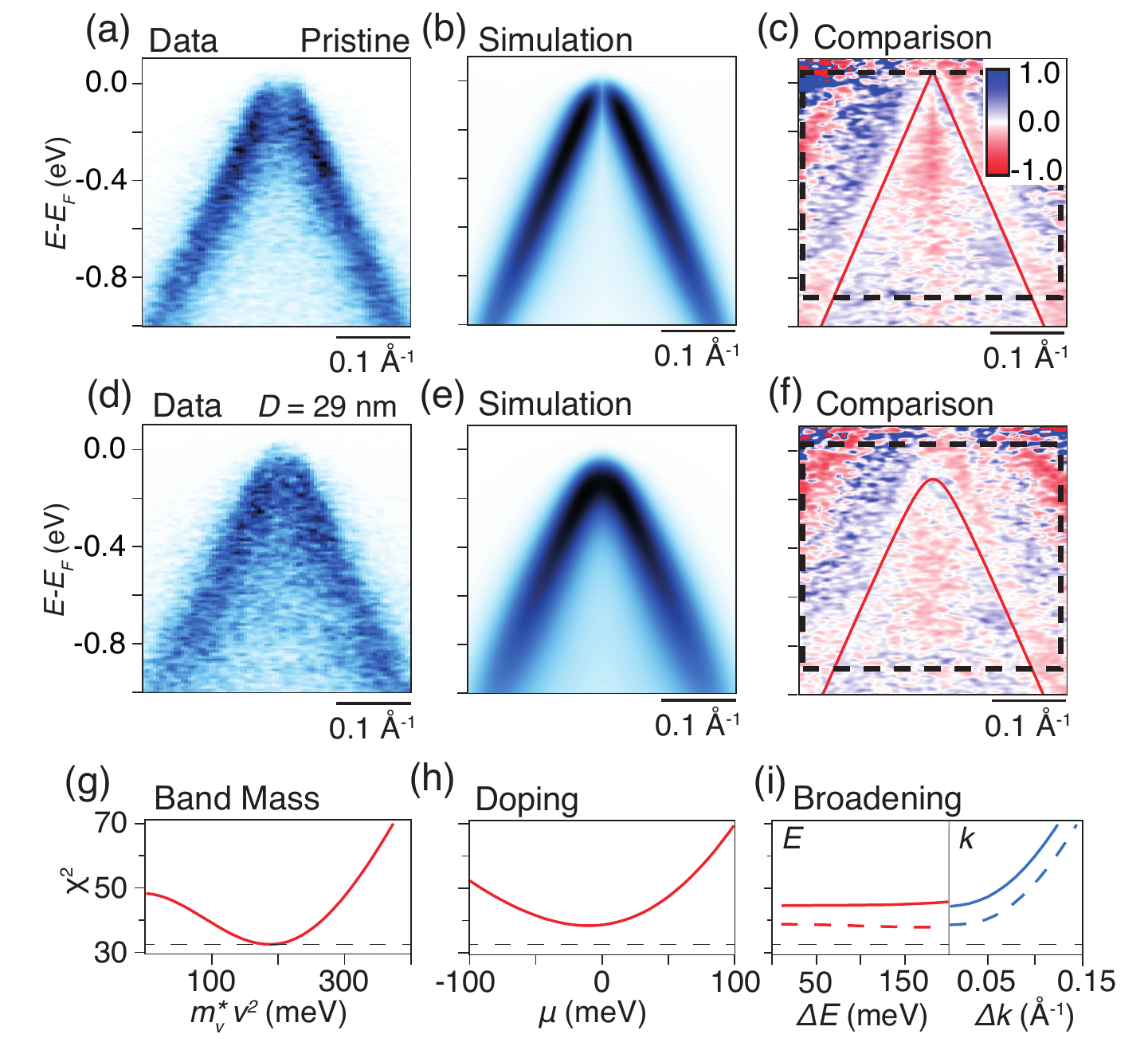}
 		\vspace*{-0.4cm}
 		\caption{Simulations of the photoemission intensity.  a) ARPES spectrum from pristine graphene.  b) Result of fitting a simulation of the intensity in (a).  c) Difference spectra between the measured and simulated data. The fitted bare bands have been overlaid in red.  d) ARPES spectrum from patterned graphene with 29 nm hole diameter.  e) Simulated photoemission intensity fitted to the data in (d).  f) Difference spectra between the measured and simulated data for patterned graphene. g) $\chi^2$-minimization for the intensity integrated within the dashed box in (f) as a function of $m_v^{\ast} v^2$.  h) $\chi^2$-minimisation of the data presented in (d) changing only the doping, showing a minimum occurring at a significantly higher $\chi^2$ value than when varying the mass.  i) $\chi^2$-minimisation of the data presented in (d) using only energy (red) or momentum (blue) broadening. Dashed blue and red lines show the same minimisation when starting from the optimised doping value determined in (h). The dashed horizontal line in (g)-(i) indicates the universal minimum of $\chi^2$, obtained by optimizing the mass term.}
 		\label{fig:S3}
 	\end{center}
 \end{figure*}

Initially, a simulation for the pristine graphene dispersion was fit, as demonstrated in Figures \ref{fig:S3}a-b. The key parameters varied were $v$, $\mu$, $\Sigma''$, and $M$, as well as a background signal with a quadratic energy- and momentum-dependence. The normalised difference spectra between the fitted and measured data is presented in Figure \ref{fig:S3}c. 
 
\begin{figure*} [h!t!]
	\begin{center}
		\includegraphics[width=1\textwidth]{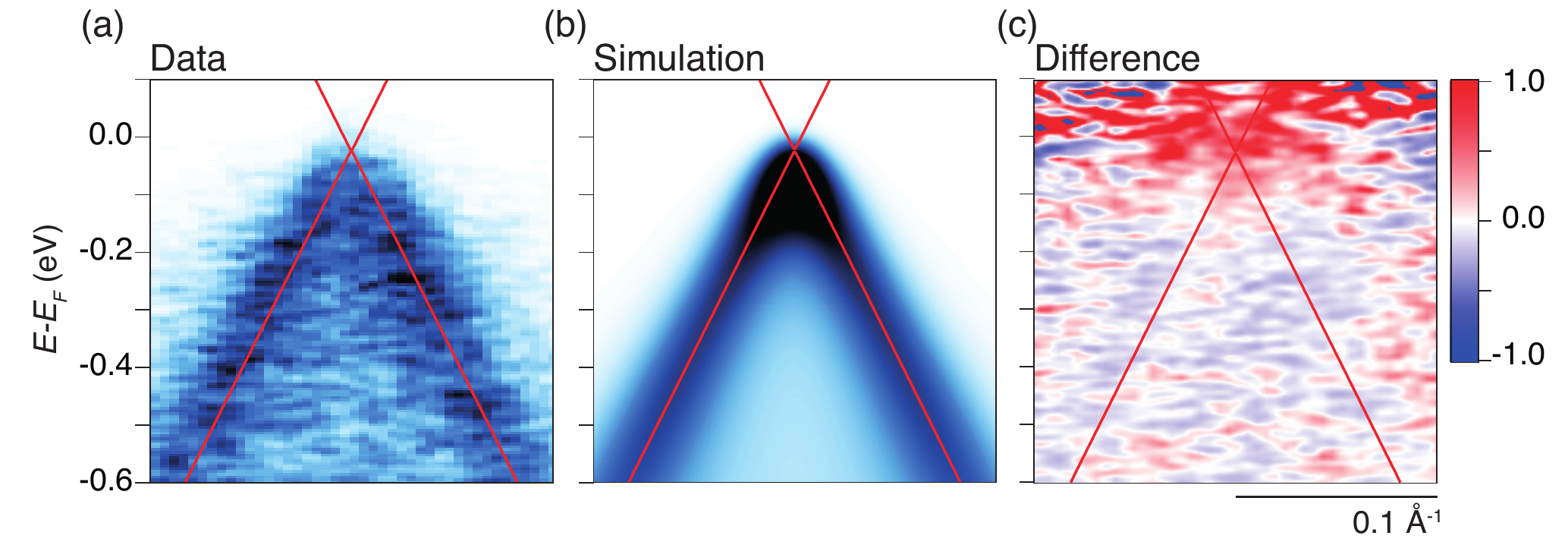}
		\caption{Simulations of photoemission intensity with optimization based on doping. a) ARPES spectrum from the same $D=29$ nm patterned graphene region as shown in Figure \ref{fig:S3}d. b) Simulated photoemission intensity generated using zero mass gap, with an optimized doping-induced shift of the Dirac point to -0.01 eV.  c) Difference spectrum between the measured and simulated graphene.}
		\label{fig:Doped}
	\end{center}
\end{figure*}
 
In order to fit the dispersion of patterned graphene, the fitted parameters from pristine graphene were used as initial conditions. $\Sigma''$, $M$ and the background were then freed and the simulation fit to the data. Finally, the mass term $m_v^\ast v^2$ was manually varied and the data re-fit with only the band intensity freed. The data and fit results for patterned graphene with $D=29$~nm are shown in Figures  \ref{fig:S3}d-f. To find the optimum mass, $\chi^2$  was calculated within the region described by a dashed box in Figure \ref{fig:S3}f. This allowed for the contributions away from the bands, and from the low intensity regions above \EF, to be minimised. Figure \ref{fig:S3}g presents an example $\chi ^2$-minimization to select the optimum mass term. 
Other terms, specifically the doping $\mu$, energy, and momentum broadening, were also run through this same procedure to ensure that our results could not be better described without a band mass. Figure \ref{fig:S3}h presents $\chi^2$ as a function of $\mu$, and shows that the minimal $\chi^2$ value is significantly higher than the minimum reached when the band mass instead was optimised. Figure \ref{fig:S3}i shows the minimisation of the energy and momentum broadening before (solid lines) and after (dashed lines) minimising the $\chi^2$ for doping. With no mass gap, but varying both $\mu$ and the Gaussian broadening, there was still a greater $\chi^2$ than for using only a mass term. This is clarified in Figure \ref{fig:Doped}, where the photoemission intensity is simulated with zero mass gap, using the doping value found by minimising $\chi^2$ in Figure \ref{fig:S3}h. The difference signal in Figure \ref{fig:Doped}c shows a clear excess of intensity around the Dirac point. \\

\newpage

\noindent\textbf{Supplementary Note 6.  Momentum broadening caused by impurity scattering}

\begin{figure*} [h!]
	\begin{center}
		\includegraphics[width=1\textwidth]{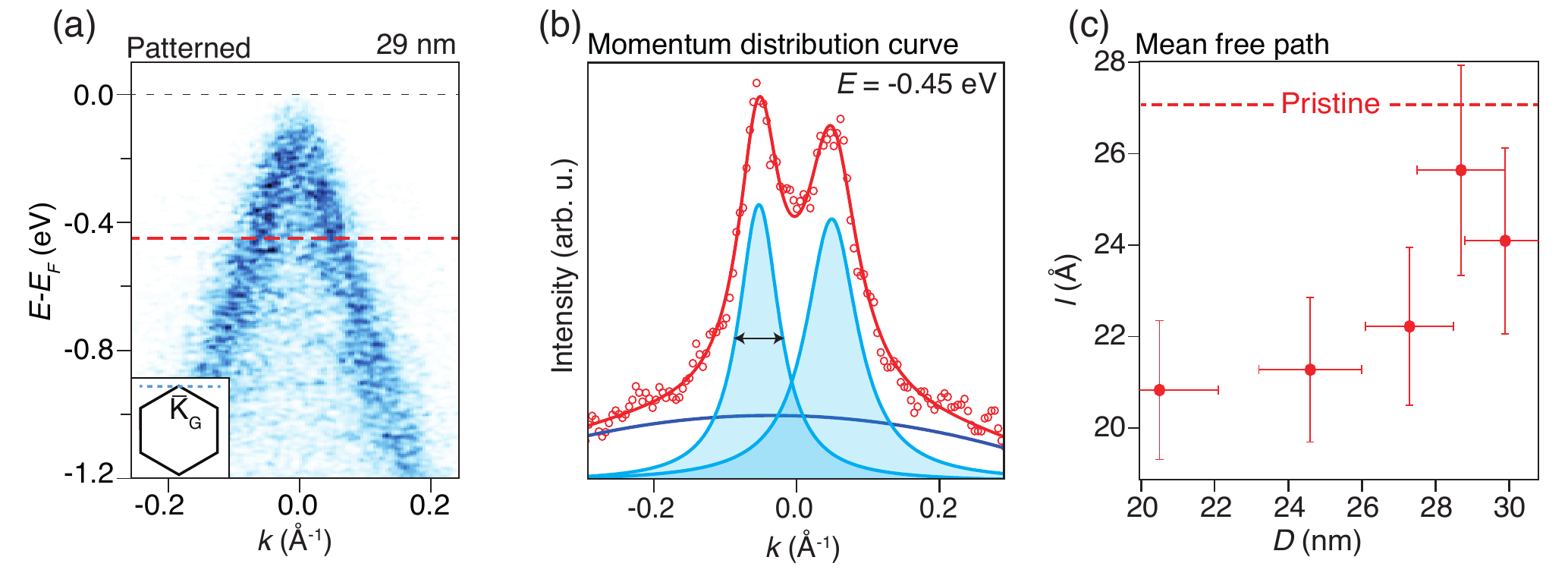}
		\vspace*{-0.8cm}
		\caption{Effect of impurity scattering on mean free path.  a) ARPES spectrum of patterned graphene with $P$ = 35 nm and $D$ = 29 nm.  b) MDC at -0.45 eV (markers) with a fit (red curve) to a function composed of two Lorentzian peaks (blue peak components) and a parabolic background (blue curve).  c) Mean free path extracted from the MDCs linewidth analysis in (b). The dashed line indicates the average mean free path for pristine graphene.}
		\label{fig:S5}
	\end{center}
\end{figure*}

The impact of impurity scattering between different patterned superstructures is determined via momentum distribution curve (MDC) analysis, as demonstrated in Figure \ref{fig:S5}. The imaginary part of the quasiparticle self energy $\Sigma''$ can be related to the MDC linewidth of graphene via the expression $\Sigma''=\hbar v \Delta_k/2$,  where $\Delta_k$ is the MDC full width at half maximum (FWHM). The quasiparticle lifetime is given by $\tau = \hbar/\Sigma''$, such that the mean free path $l = \tau v$ is given in terms of the MDC FWHM via $l = 2/\Delta_k$ for graphene. The measured MDC linewidth thereby encodes the various quasiparticle scattering mechanisms of the system. Impurity scattering contributes to these measurements via an energy- and momentum-independent broadening term \cite{Bostwick2007}. 

In order to demonstrate the analysis, we plot an ARPES dispersion around $\bar{\mathrm{K}}_{\mathrm{G}}$ from patterned graphene in Figure \ref{fig:S5}a and extract an MDC at -0.45 eV  in Figure \ref{fig:S5}b.  A double Lorentzian peak on a parabolic background provides an excellent fit of the MDC, as shown in Figure \ref{fig:S5}b. We determine $\Delta_k$ as an average of the FWHM of the two Lorentzian peaks and plot the resulting mean free path for different superstructures in Figure \ref{fig:S5}c. These data correspond to the superstructures with the same hole diameter and periodicity as those shown in the main text. The average value of the mean free path for pristine graphene is marked by a dashed line.  As expected from the extra sample processing, the patterned regions exhibit a shorter mean free path than the pristine graphene.  Interestingly,  the mean free path is longer for superstructures with larger hole diameters,  indicating that impurity scattering is weaker for these particular regions.  Based on this observation we can rule out that the increasing effective mass and shift of EDC leading edge with increasing hole diameter are related to a simultaneous increase in impurity scattering due to e.g. edge-disorder around holes or lattice disorder caused by the lithographic etching.  \\

\newpage

\noindent\textbf{Supplementary Note 7.  G$\mathbf{_0}$W$\mathbf{_0}$ band structures and effective mass fits}

In this section, we provide G$_0$W$_0$ band structures and effective mass fits. The single-shot G$_0$W$_0$ calculation is made on top of a PBE-DFT calculation that yields a band gap of $E_{\text{PBE}} = 2.04$ eV and effective masses $m_c^* = 0.46 m_e$ and $m_v^* = 0.79 m_e$ for all jellium charge densities. In fact, the only part of the DFT results sensitive to the jellium charge is the Fermi level. The obtained PBE band gap is somewhat smaller than the LDA band gap of $E_{\text{LDA}} = 2.34$ eV and HSE0 band gap of $E_{\text{HSE0}} = 3.2$ eV found by Du \textit{et al.} for the same structure \cite{Aijun:2010}. The quasiparticle effective masses are fitted from bands inside a circular region around K, consisting of 31 $k$-points. We fit to a quasi-relativistic dispersion on the form $E(k) = \alpha + \beta \sqrt{1 + \frac{k^2}{\gamma + \eta k^2}}$, yielding an effective mass $m^* = \frac{\gamma \hbar^2}{\beta}$. This function was chosen to capture the relativistic nature of graphene, while the $\eta k^2$ term serves to capture band flattening due to non-hyperbolicity away from the Dirac point. The low $k$-point density applied in G$_0$W$_0$ calculations is dictated by computational demands of the structure, and could potentially lead to unreliable fits.

\begin{figure*} [h!]
	\begin{center}
		\includegraphics[width=0.7\textwidth]{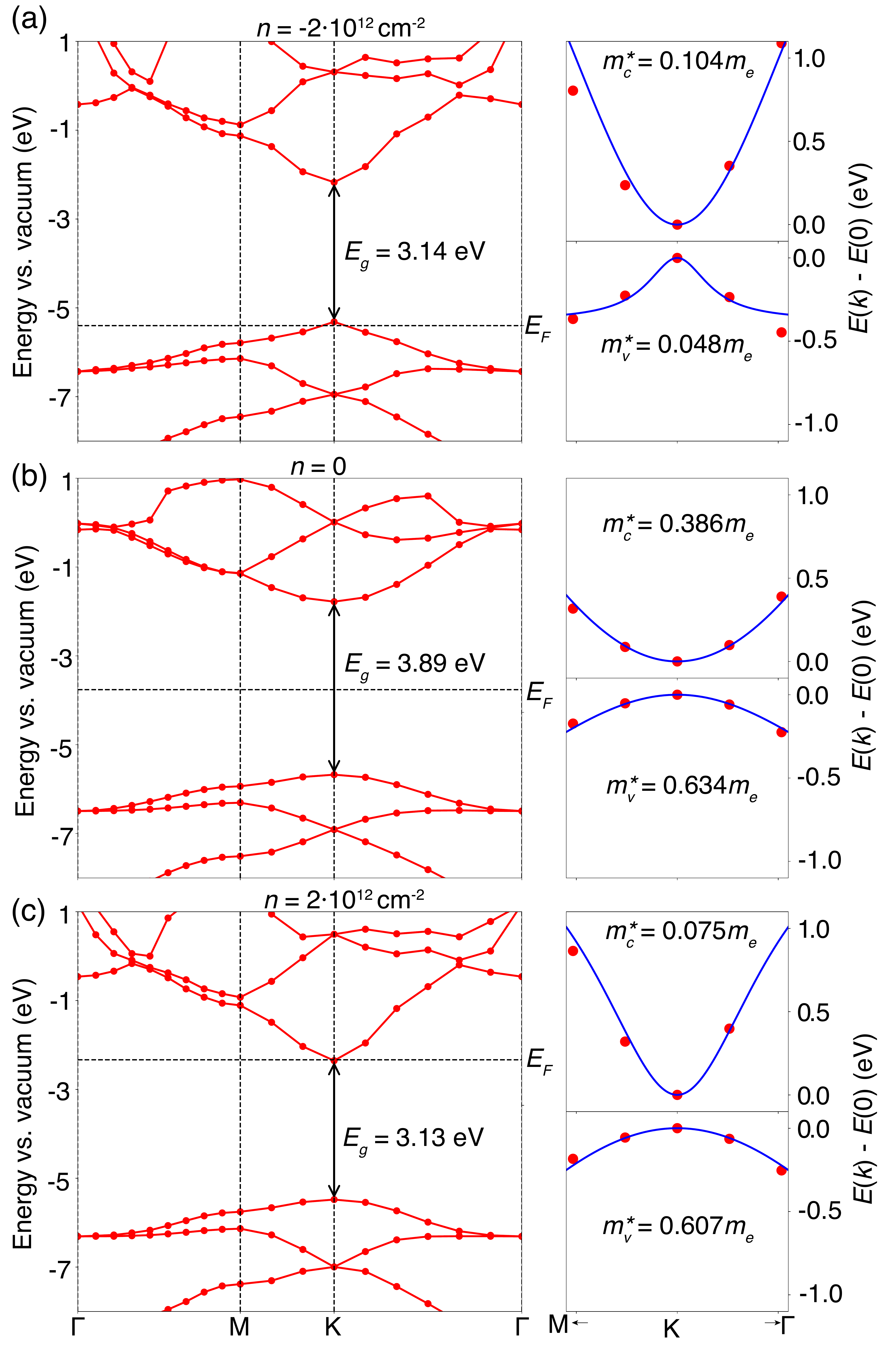}
		\vspace*{-0.3cm}
		\caption{G$_0$W$_0$ band structure (left), effective mass fit of the conduction band (top-right), and effective mass fit of the valence band (bottom-right), for the structure shown in Figure 4c of the main text, with a jellium charge density of (a)-(c): $\{-2, 0, 2\} \cdot 10^{12} \, \text{cm}^{-2}$.}
		\label{fig:S6}
	\end{center}
\end{figure*}

We have, however, verified that purely parabolic fits using only the nearest neighboring $k$-points to K yield effective masses similar to our full quasi-relativistic fits. The deviation between the two approaches is less than $0.02 m_e$ for the larger masses, indicating that these results are robust. In particular, the two approaches yield nearly identical hole masses for moderate and large $n$-doping.
In the case of smaller effective masses, the deviation becomes significantly larger, and the results are less reliable due to the low $k$-point density. The curves in Figure \ref{fig:S6} show the full G$_0$W$_0$ band structures, as well as the fitted quasi-relativistic dispersions along the M-K-$\Gamma$ line in the vicinity of K. The robust valence band fits provide the hole masses shown in Figure 4c of the main text.\\

\newpage

\noindent\textbf{Supplementary Note 8.  Renormalization of band gap and effective masses}

In this section, we provide a simplified model for the density dependence of band gap renormalization in two dimensions and discuss effective mass renormalization in terms of a two-band quasi-relativistic model. The starting point is the screened RPA (random phase approximation) expression for the electron and hole self-energies $\Sigma_{\alpha}$ with $\alpha =c,v$ in the plasmon-pole approximation \cite{thuselt,SchmittRink}. For a two-dimensional (2D) semiconductor with effective Bohr radius $a_0^*$ and areal carrier density $n$, this approach is valid in the perturbative regime $r_s^{-1}\approx (a_0^*)^2|n|\gg 1$, in which Coulomb interactions are weak compared to kinetic energy \cite{Hawrylek}. This implies that the weak doping regime is not covered by the model. Rather, its virtue is the simplicity, which allows for analytical estimates of band renormalization due to carrier effects. We expand slightly on the usual approach \cite{thuselt,SchmittRink} by including nonlocal ($q$-dependent) screening and write
\begin{equation}
	{{\Sigma }_{\alpha }}(\vec{k})=-\frac{1}{\Omega }\sum\limits_{{\vec{q}}}{{{V}_{{\vec{q}}}}f_{\vec{k}-\vec{q}}^{\alpha }}-\frac{1}{2\Omega }\sum\limits_{{\vec{q}}}{{{V}_{{\vec{q}}}}\frac{\omega _{p}^{2}}{{{\omega }_{{\vec{q}}}}}\left\{ \frac{1-f_{\vec{k}-\vec{q}}^{\alpha }-{{g}_{{\vec{q}}}}}{E_{{\vec{k}}}^{\alpha }-E_{\vec{k}-\vec{q}}^{\alpha }-\hbar {{\omega }_{{\vec{q}}}}}+\frac{f_{\vec{k}-\vec{q}}^{\alpha }+{{g}_{{\vec{q}}}}}{E_{{\vec{k}}}^{\alpha }-E_{\vec{k}-\vec{q}}^{\alpha }+\hbar {{\omega }_{{\vec{q}}}}} \right\}}.\nonumber
\end{equation}	
Here, $\Omega$ is a normalization volume or area in 3D and 2D, respectively, $\omega_{p}$ is the plasmon frequency and $\omega _{{\vec{q}}}^{2}=\omega _{p}^{2}+{{(\hbar {{q}^{2}}/2\mu )}^{2}}$ with $\mu =2m_c^*m_v^*/(m_c^*+m_v^*)$ twice the reduced electron-hole mass. Also, $E_{{\vec{k}}}^{\alpha }={{\hbar }^{2}}{{k}^{2}}/2{{m}_{\alpha }^*}$ and $f_{{\vec{k}}}^{\alpha }=1/(\exp [(E_{\vec{k}}^{\alpha }-E_F)/{{k}_{B}}T]+1)$, with chemical potential $E_F$, and ${{g}_{{\vec{q}}}}=1/(\exp [\hbar {{\omega }_{{\vec{q}}}}/{{k}_{B}}T]-1)$. The layered 2D structure implies that the bare Coulomb interaction ${{V}_{{\vec{q}}}}$ is of the Keldysh form \cite{Keldysh} with nonlocal screening through a $q$-dependent dielectric constant $\varepsilon(q)$. In its linearized form, $\varepsilon(q)=\bar{\varepsilon}+r_0 q$ with $\bar{\varepsilon}$ the average of dielectric constants above and below the structure and $r_0$ the characteristic screening length of the 2D semiconductor. The effective Bohr radius may be estimated as $a_0^*\approx {\bar{\varepsilon}}^2(2m_e/\mu)\times 0.53\text{\AA}\approx 10 ^{-6}$ $\text{cm}$ assuming $\mu=0.1 m_e$ (see main text) and ${\bar{\varepsilon}}^2\approx 10$. This implies a lower density limit of $\sim 10 ^{12}$ $\text{cm}^{-2}$ for the RPA model. We assume that the plasmon pole is determined by the local value $\bar{\varepsilon}$ so that
\begin{equation}
{{V}_{{\vec{q}}}}=\frac{{{e}^{2}}}{2{\varepsilon }_{0}\varepsilon(q) q},\quad \omega _{p}^{2}=q\Omega _{p}^{2},\quad \Omega _{p}^{2}=\frac{{{e}^{2}}|n|}{{\varepsilon }_{0}\bar{\varepsilon}\mu }.\nonumber
\end{equation}
We now take the low temperature limit such that ${{g}_{{\vec{q}}}}\approx 0$ and rewrite as
\begin{equation}
	{{\Sigma }_{\alpha }}(\vec{k})=\frac{1}{\Omega }\sum\limits_{{\vec{q}}}{\frac{\omega _{p}^{2}}{{{\omega }_{{\vec{q}}}}}\frac{{{V}_{{\vec{q}}}}}{E_{\vec{k}-\vec{q}}^{\alpha }-E_{{\vec{k}}}^{\alpha }+\hbar {{\omega }_{{\vec{q}}}}}-}\frac{1}{\Omega }\sum\limits_{{\vec{q}}}{f_{\vec{k}-\vec{q}}^{\alpha }{{V}_{{\vec{q}}}}\left\{ 1-\frac{\hbar \omega _{p}^{2}}{{{(E_{{\vec{k}}}^{\alpha }-E_{\vec{k}-\vec{q}}^{\alpha })}^{2}}-{{(\hbar {{\omega }_{{\vec{q}}}})}^{2}}} \right\}}.\nonumber
\end{equation}
The first and second terms are the Coulomb hole and screened exchange term, respectively, and we retain only the dominant Coulomb hole \cite{thuselt,SchmittRink}. The results simplify greatly when expressed through the scaled 2D plasmon frequency $\nu\equiv (2\mu {{\Omega }_{p}}/\hbar)^{2/3}$ that varies with density as $\nu\propto{|n|}^{1/3}$.  Accordingly, we introduce $t={(q/\nu)}^{3/2}$, $\rho =r_0\nu/\bar{\varepsilon}$ and $\kappa =k/\nu$ and then get
\begin{equation}
\Sigma _{\alpha }(\vec{k})=\frac{{{e}^{2}}\nu}{6{{\pi }^{2}}{{\varepsilon }_{0}\bar{\varepsilon }}\hbar }\int\limits_{0}^{\infty }{\int\limits_{0}^{\pi }{\frac{d\theta dt}{{{(1+{{t}^{2}})}^{1/2}}(1+\rho t^{2/3})[\tfrac{\mu }{{{m}_{\alpha }^*}}(t-2\kappa {{t}^{1/3}}\cos \theta )+{{(1+{{t}^{2}})}^{1/2}}]{{t}^{1/3}}}}}.\nonumber
\end{equation}
Restricting to the band gap location $k=0$ and assuming $\mu ={{m}_{\alpha }^*}$ for simplicity, we finally find the self-energy $\Sigma _{\alpha }\equiv \Sigma _{\alpha }(0)$ to first order in $r_0$
\begin{equation}
\Sigma _{\alpha }=\frac{{{e}^{2}\nu}}{3\pi{\varepsilon}_{0}\bar{\varepsilon}\hbar }\left\{\frac{3\sqrt{\pi }\Gamma (\tfrac{2}{3})}{\Gamma (\tfrac{1}{6})}-\frac{r_0\nu }{\bar{\varepsilon}}\frac{\Gamma (\tfrac{1}{3})\Gamma (\tfrac{1}{6})}{8\sqrt\pi}\right\}.\nonumber 
\end{equation}
This demonstrates that the dominant self-energy varies as ${|n|}^{1/3}$ in 2D with corrections of the form ${|n|}^{2/3}$ in cases of pronounced nonlocal screening. A similar calculation leads to a ${|n|}^{1/4}$ dependence for the local term in 3D systems \cite{thuselt,SchmittRink}. This qualitative difference reflects the constant plasmon frequency $\omega _{p}^{2}=2{{e}^{2}}|n|/(\varepsilon {{\varepsilon }_{0}}\mu)$ characteristic of 3D systems with volume density $n$, in contrast to the linear $q$-dependence in 2D. In the above form, the assumption $\mu ={{m}_{\alpha }^*}$ implies identical self-energies for the two bands. In reality, the  $\mu/{{m}_{\alpha }^*}$ factor in the denominator of the full expression demonstrates that the heavier band of the two contributes more to the total renormalization.

The renormalized band gap is $E_g(n)=E_g^{(0)}(1-\beta _c-\beta _v)$ with $\beta_\alpha\equiv\Sigma _{\alpha }/E_g^{(0)}$. As demonstrated in Figure 4c of the main paper, outside the weak-doping regime the full G$_0$W$_0$ data agree reasonably with a symmetric ${|n|}^{1/3}$ behaviour for both signs of the doping. Somewhat surprisingly, the effective hole masses seem to follow a similar density dependence for moderate and large $n$-doping, c.f. Figure 4c in the main text. A rationale for this finding can be provided by a two-band Dirac model, however. To this end, we consider a generic two-band Dirac model spanned by two pseudo-spin states
\begin{equation}
	H=\left( \begin{matrix}
   {{\alpha }_{c}} & \hbar v(k_x-i k_y)  \\
   \hbar v(k_x+i k_y) & {{\alpha }_{v}} \nonumber 
\end{matrix} \right)
\end{equation}
with Fermi velocity $v$ and doping-dependent "on-site" elements $\alpha_{v,c}$. The pseudo-spin states represent the valence and conduction band eigenstates at the band gap $\vec{k}=0$, which become coupled by any finite $k$-vector through the off-diagonal terms. The band gap is $\alpha_c-\alpha_v$. Moreover, if the pseudo-spin states are assumed orthogonal, identical electron and hole masses $m_c^*=m_v^*=(\alpha_c-\alpha_v)/(2v^2)$ result. This is clearly in contradiction to the G$_0$W$_0$ data and we, therefore, assume non-orthogonal pseudo-spin states represented by an overlap matrix 
\begin{equation}
	S=\left( \begin{matrix}
   1 & \hbar \eta v(k_x-i k_y)  \\   \hbar \eta v(k_x+i k_y) &1 \nonumber 
\end{matrix} \right),
\end{equation}
where $\eta$ is a measure of non-orthogonality that can be estimated as $\eta=s/t$ with $s$ and $t$ overlap and hopping matrix element between pseudo-spin states, respectively. Solving the secular equation $|H-E_{\alpha}S|=0$ yields quasi-relativistic hyperbolic dispersions that, upon expanding to second order in $k$, become
\begin{equation}
{{E}_{c}}(k)={{\alpha }_{c}}+\frac{{{\hbar }^{2}}{{v}^{2}}{{(1-\eta{{\alpha }_{c}} )}^{2}}}{{{\alpha }_{c}}-{{\alpha }_{v}}}{{k}^{2}},\quad {{E}_{v}}(k)={{\alpha }_{v}}-\frac{{{\hbar }^{2}}{{v}^{2}}{{(1-\eta{{\alpha }_{v}} )}^{2}}}{{{\alpha }_{c}}-{{\alpha }_{v}}}{{k}^{2}}.\nonumber  
\end{equation}
This demonstrates that  $\eta\neq 0$ is a requirement for different electron and hole masses in the two-band Dirac model. While the on-site difference is bound by  $\alpha_c-\alpha_v=E_g^{(0)}(1-\beta _c-\beta _v)$, a particularly simple result is found if the energy zero-point is selected such that the average  $(\alpha_c+\alpha_v)/2$ is set equal to $\eta^{-1}$ in the intrinsic material. In the case $\alpha_c=\eta^{-1}+(1/2-\beta_c)E_g^{(0)}$ and  $\alpha_v=\eta^{-1}-(1/2-\beta_v)E_g^{(0)}$, the effective masses become
\begin{align}
m_c^*=m_0\frac{1-\beta _c-\beta _v}{(1-2\beta _c)^2}\approx m_0(1+3\beta _c-\beta _v),\nonumber \\ 
m_v^*=m_0\frac{1-\beta _c-\beta _v}{(1-2\beta _v)^2}\approx m_0(1+3\beta _v-\beta _c),\nonumber 
\end{align}
where we introduced the bare (un-renormalized) mass as $m_0=2/(v^2\eta^2 E_g^{(0)})$ and assumed $\Sigma _{\alpha } \ll E_g^{(0)}$. Thus, in this scenario, bare electron and hole masses are identical but  differ in their renormalization. Importantly, the renormalization factor follows exactly the same form as the band gap, provided the ratio $\beta_c/\beta_v$ is only weakly dependent on carrier density. Moreover, if $\beta_v>3\beta_c$, electron and hole masses will decrease and increase with density, respectively. This appears to be the case for $n$-doping, while the opposite may be true for $p$-doping. This sensitivity to dopant type is captured in the $\text{sign}(n)$ factor included in the fit in Figure 4c of the main text.

\end{document}